\documentclass[12pt]{article}
\usepackage{pgfplots}
\usetikzlibrary{decorations.pathmorphing}

\tikzset{snake it/.style={decorate, decoration=snake}}
\pgfplotsset{compat=1.10}
\usepgfplotslibrary{fillbetween}
\usetikzlibrary{patterns}
\usepackage{draft} 

\usepackage{hyperref}
\usepackage{graphicx,color,subfigure}
\usepackage{cite}
\usepackage{mciteplus}
\usepackage{skak}
\usepackage{xcolor}
\usepackage{empheq}
\usepackage{tikz}
\DeclareFontFamily{OT1}{pzc}{}
\DeclareFontShape{OT1}{pzc}{m}{it}{<-> s * [1.10] pzcmi7t}{}
\DeclareMathAlphabet{\mathpzc}{OT1}{pzc}{m}{it}

\def\be#1\ee{\begin{align}#1\end{align}}



\begin{document}

\unitlength = .8mm

\begin{titlepage}

\begin{center}

\hfill \\
\hfill \\
\vskip 1cm

\title{\Huge Interpolating single and double trace $T\bar T$  \vskip 0.1
cm  by moving LST into the bulk}

\author{Jinwei Chu$^a$, Amit Giveon$^b$ and David Kutasov$^a$}
\address{$^a$Leinweber Institute for Theoretical Physics, Enrico Fermi Institute,\\ 
and Department of Physics, University of Chicago, Chicago IL 60637, USA\\
\vspace{2mm}
$^b$Racah Institute of Physics, The Hebrew University, Jerusalem, 91904, Israel}
\vskip 1cm

\email{jinweichu@uchicago.edu,giveon@mail.huji.ac.il,dkutasov@uchicago.edu}

\end{center}

\abstract{We describe a large class of theories that continuously interpolate between single and double trace $T\bar T$ deformations of AdS$_3$/CFT$_2$. It is obtained by starting with the spacetime ${\cal M}_3$, which interpolates between AdS$_3$ in the IR and a linear dilaton spacetime in the UV, and imposing a UV cutoff on the radial direction. When the cutoff is deep in the AdS$_3$ region of ${\cal M}_3$, this leads to double trace $T\bar T$. When it goes to infinity, we recover the single trace deformed theory.  We present a large class of possible boundary conditions on the dilaton, and compute the resulting black hole spectrum in a few examples. In all these constructions, to leading order in the low energy expansion, the resulting spectrum is that of a single $+$ double trace deformed CFT, but at higher orders the different theories have different spectra. In one of the examples that we analyze, the full theory is single $+$ double trace $T\bar T$ deformed CFT. Another gives rise to an interesting spectrum, that consists of an infinite set of energy bands.

}

\vfill

\end{titlepage}

\eject

\begingroup
\hypersetup{linkcolor=black}
\tableofcontents
\endgroup

\vfill
\eject

\vfill
\eject

\section{Introduction}
\label{intro}

The AdS/CFT correspondence is one of the most successful paradigms in theoretical physics of recent decades (see e.g. \cite{Aharony:1999ti} for a review). It posits an equivalence between a quantum theory of gravity in anti-de Sitter spacetime, and a non-gravitational theory -- a Conformal Field Theory that lives in one lower dimension, on the boundary of AdS. It is widely expected that this correspondence is a special case of a more general holographic duality between quantum theories of gravity and non-gravitational theories. Understanding this more general duality is an important open problem.

There are two different approaches to understanding holography more broadly. One is to improve our understanding of AdS/CFT, and this has been a very fruitful direction of research in recent years. Another is to push holography  beyond AdS/CFT. There has been a lot of work on gravity in spacetimes that are asymptotically AdS near the boundary, and their duals, which are QFT's that approach an RG fixed point in the UV (see e.g. \cite{Skenderis:2002wp} for a review). Since these deformations do not change the boundary structure of the theory, they constitute relatively straightforward generalizations of the original AdS/CFT correspondence. 

Another class of spacetimes that has been studied from the point of view of holographic duality is asymptotically linear dilaton spacetimes. They appear in the context of $1+1$ dimensional string theory \cite{Klebanov:1991qa,Ginsparg:1993is}, the physics of fivebranes (known as Little String Theory) \cite{Aharony:1998ub,Aharony:1999ks,Kutasov:2001uf}, and in the case of AdS$_3$, which will be particularly relevant for us here, in single trace $T\bar T$ deformed CFT \cite{Giveon:2017nie}. The last example provides an interesting blueprint of a way to proceed in generalizing the AdS/CFT correspondence. It starts with that correspondence, and deforms the boundary theory by adding to the Lagrangian a particular irrelevant operator, that has the interesting property that its effect on the theory can be controlled, both in the bulk and on the boundary. 

Another example of a UV deformation of the AdS$_3$/CFT$_2$ correspondence that follows this blueprint is three dimensional gravity in a spherical cavity in AdS$_3$. It was argued in \cite{McGough:2016lol} that on the boundary this corresponds to a (double trace) $T\bar T$ deformation with negative coupling of the original CFT. It has a finite entropy, and thus is an example of theories with this property, that are of more general interest. 

The success of single and double trace $T\bar T$ deformed AdS$_3$/CFT$_2$ motivates exploring generalizations. In this paper, we take a step in this direction. The basic idea is to combine the two deformations. We start with the bulk spacetime that describes single trace $T\bar T$ deformed CFT, which interpolates between linear dilaton spacetime in the UV and AdS$_3$ in the IR, and impose a UV cutoff by placing it in a radial cavity. We find that unlike the AdS$_3$ case, there are many options for imposing boundary conditions at the cavity wall, and they lead to interesting theories, some with rather exotic properties. 

A common theme of all the constructions is that they provide a continuous interpolation between single and double trace deformed CFT. We provide a bulk analysis of these interpolating theories, and some of their properties, leaving a better understanding of the corresponding boundary theories to future work. 

The structure of this paper is the following. In section \ref{sectwo}, we review some results on single trace $T\bar T$ deformed CFT. In section \ref{secthree}, we review AdS$_3$ gravity in a spherical cavity. We use a canonical ensemble approach to the theory to derive the deformed energy of BTZ black holes, in preparation to our analysis of the combined system. 

In section \ref{secfour}, we impose a spherical cutoff on the bulk spacetime corresponding to single trace $T\bar T$ deformed CFT. We show that there are many possible boundary conditions one can impose, and study some examples of the resulting theories. In particular, we show that one choice of boundary conditions leads to a theory that can be described as the original CFT$_2$ deformed by single and double trace $T\bar T$ perturbations. Another choice leads to a theory with a rather exotic spectrum, that contains an infinite number of energy bands. 

Section \ref{secfive} contains some discussion of our results.

\section{Positive single-trace $T\bar T$}
\label{sectwo}

Single trace $T\bar{T}$ holography \cite{Giveon:2017nie} is obtained by studying type II string theory in the near-horizon geometry of a system of Neveu-Schwarz fivebranes (which we will refer to as $NS5$-branes), in the presence of fundamental strings (which we will refer to as $F1$'s) at the core of the fivebrane geometry. As is well known \cite{Aharony:1999ti}, taking the near-horizon limit of the $F1$'s as well leads to an AdS$_3$ background, that corresponds via the AdS/CFT correspondence to a two dimensional CFT, often referred to as the boundary or spacetime CFT. Much is known about this CFT, but some of its most important features  remain enigmatic. 

If we do not take the near-horizon limit of the $F1$'s, we get a three dimensional background that interpolates between AdS$_3$ in the infrared and a linear dilaton background in the UV. This background, which we will refer to as ${\cal M}_3$, can be understood from the boundary point of view as an irrelevant perturbation of the above boundary CFT, or as a particular $1+1$ dimensional vacuum of a $5+1$ dimensional non-local theory known as Little String Theory (or LST) \cite{Aharony:1998ub,Aharony:1999ks,Kutasov:2001uf}. As we explain below, the detailed structure of the resulting boundary theory is the origin of the name {\it single trace $T\bar T$ deformed CFT} used in the literature. 

In the rest of this section we review the construction of single trace $T\bar T$ deformed CFT and its bulk dual. As discussed in section \ref{intro}, it serves as one of the two pillars on which we will build below. The other will be reviewed in section \ref{secthree}. We will use results from \cite{Chakraborty:2020swe,Chakraborty:2023zdd}, which can be consulted for more details and references to prior literature.

\subsection{The $F1$ --  $NS5$ system}
\label{twoone}

Our starting point is a $4+1$ dimensional vacuum of type II string theory, $\mathbb{R}^{1,4}\times\mathbb{S}^1\times \mathbb{T}^4$. The circle in the geometry, which we will label by $x$, plays an important role in the construction. We denote its radius by $R$, 
\be\label{xR}
x\sim x+2\pi R~.
\ee
We add to the background $k$ $NS5$-branes wrapping $\mathbb{S}^1\times \mathbb{T}^4$, and $p$ $F1$'s that wrap the $\mathbb{S}^1$, and carry $n$ units of momentum on the circle.  The branes deform the metric and dilaton in their vicinity to
\be\label{ds}
ds^2={1\over f_1}\left[-{f\over f_n}dt^2+f_n\left(dx-{r_0^2\sinh 2\alpha_n\over 2f_nr^2}dt\right)^2\right]+f_5\left({dr^2\over f}+r^2d\Omega_3^2\right)+\sum_{i=1}^4dx_i^2~,
\ee
and
\be\label{dil}
e^{2\Phi}=g^2{f_5\over f_1}~,
\ee
respectively. Here,
\be\label{fs}
f=1-{r_0^2\over r^2}~,\qquad f_{1,5,n}=1+{r_{1,5,n}^2\over r^2}~,
\ee
with
\be\label{alphas}
r_{1,5,n}^2=r_0^2\sinh^2\alpha_{1,5,n}~,
\ee
and
\be\label{kp}
\sinh 2\alpha_1={2\alpha'pg^2\over vr_0^2}~,\qquad\sinh 2\alpha_5={2\alpha'k\over r_0^2}~,\qquad\sinh 2\alpha_n={2\alpha'^2g^2n\over vR^2r_0^2}~.
\ee
A few comments about these equations:
\begin{itemize}
\item $(r,\Omega_3)$ are spherical coordinates on the $\mathbb{R}^4$ transverse to the branes. 
\item $r_0$ is the Schwarzschild radius, associated with the fact that we took the fivebranes to be non-extremal.
\item $v$ is the dimensionless volume of the $\mathbb{T}^4$, $V_{\mathbb T^4}=v(2\pi)^4\alpha'^2$. $g$ is the string coupling at infinity, \eqref{dil}.
\item $(k,p,n)$ in \eqref{kp} are the numbers of $NS5$-branes, $F1$'s and units of momentum on the $x$-circle, respectively. 
\item There is also a $B$-field that we did not write, since it will not play a role below. It can be found in the references mentioned above. 
 \end{itemize}
The ADM mass of the brane system is
\be\label{adm}
M_{\rm ADM}={Rvr_0^2\over 2\alpha'^2g^2}(\cosh 2\alpha_5+\cosh 2\alpha_1+\cosh 2\alpha_n)~.
\ee
Subtracting from it the  ground-state energy of the system, 
\be\label{ext}
E_{\rm ext}={kRv\over\alpha'g^2}+{pR\over\alpha'}={Rvr_0^2\over 2\alpha'^2g^2}(\sinh 2\alpha_5+\sinh 2\alpha_1)~,
\ee
we get the energy above the ground state,
\be\label{E}
E\equiv M_{\rm ADM}-E_{\rm ext}={Rvr_0^2\over 2\alpha'^2g^2}\left(e^{-2\alpha_5}+e^{-2\alpha_1}+\cosh 2\alpha_n\right)~.
\ee 
The Bekenstein-Hawking (BH) entropy is
\be\label{S}
S={2\pi Rvr_0^3\over\alpha'^2g^2}\cosh\alpha_5\cosh\alpha_1\cosh\alpha_n~,
\ee
and the inverse Hawking temperature is
\be\label{betaa}
\beta=2\pi r_0\cosh\alpha_5\cosh\alpha_1\cosh\alpha_n~.
\ee
The background described in this subsection includes the region far from the branes. To study it using holography we next take the near-horizon limit of the $NS5$-branes.

\subsection{The near-$NS5$ limit}
\label{twotwo}

To focus on the dynamics of the $NS5$-branes, it is useful to consider $r$, $r_0$ and $r_n$ of order the asymptotic string coupling $g$, and take $g\to 0$ \cite{Callan:1991at,Maldacena:1997cg,Aharony:1998ub} . To implement this limit, we take $r\to gr$, $r_0\to gr_0$, $r_{1,n}\to gr_{1,n}$, and consider the limit $g\to 0$, with the rescaled $r$, $r_0$ and $r_{1,n}$ held fixed.  In this limit, 
the metric and dilaton in \eqref{ds} and \eqref{dil} take the form
\be\label{dsd}
ds^2={1\over f_1}\left[-{f\over f_n}dt^2+f_n\left(dx-{r_0^2\sinh 2\alpha_n\over 2f_nr^2}dt\right)^2\right]
+{r_5^2dr^2\over fr^2}+r_5^2d\Omega_3^2+\sum_{i=1}^4dx_i^2~,
\ee
and
\be\label{dild}
e^{2\Phi}={r_5^2\over f_1r^2}~,
\ee
where
\be\label{k}
r_5=\sqrt{\alpha'k}=\sqrt k l_s~,
\ee
and $f,f_1,f_n$ are given by \eqref{fs}, \eqref{alphas}, 
but now with
\be\label{kpd}
\sinh 2\alpha_1={2\alpha'p\over vr_0^2}~,\qquad\sinh 2\alpha_n={2\alpha'^2n\over vR^2r_0^2}~.
\ee
In the extremal case, $r_0=n=0$, the background \eqref{dsd} -- \eqref{kpd} can be described as ${\cal M}_3\times \mathbb{S}^3\times \mathbb{T}^4$ \cite{Giveon:2017nie,Chakraborty:2020swe}.
The compact space $\mathbb{S}^3\times \mathbb{T}^4$ will not play a role in this paper, since we will not attempt to identify the microstates of the black holes that we will study. Thus, we will usually omit it.

\begin{figure}
	\centering
\includegraphics[scale=0.2]{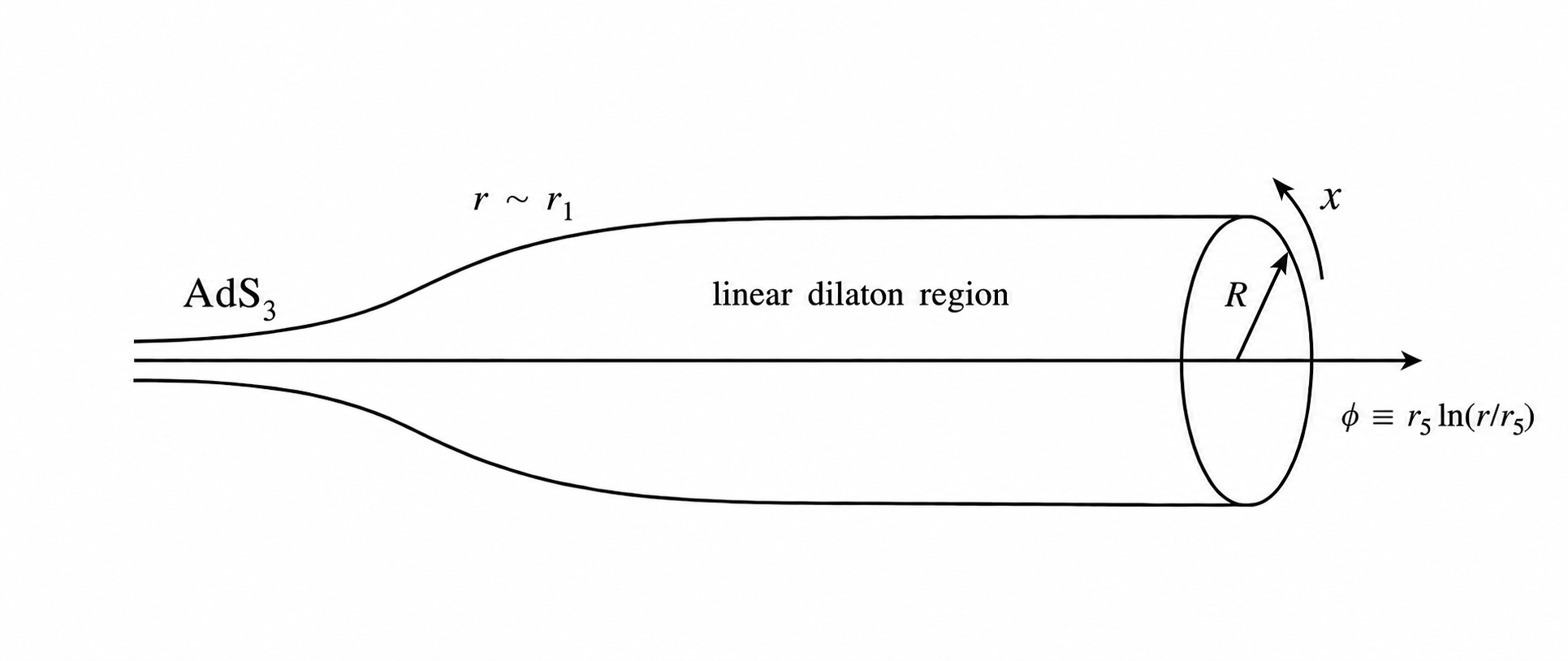}
\caption{\label{M3fig} The $2+1$ dimensional background ${\cal M}_3$ asymptotes to $\mathbb{R}_t\times\mathbb{R}_\phi \times\mathbb{S}^1$ with linear dilaton in $\phi$ as $\phi\to+\infty$ and to AdS$_3$ as $\phi\to-\infty$. The time coordinate $t$ is suppressed.}
\end{figure}

The three-dimensional spacetime labeled by $(t,x,r)$,  ${\cal M}_3$, interpolates between an asymptotically flat spacetime 
with a linear dilaton,  $\Phi\sim -\phi/r_5\equiv -\log(r/r_5$), at $r\gg r_1$,  
and AdS$_3$ (more precisely, massless BTZ) with radius $r_5$ for $r\ll r_1$.
See figure \ref{M3fig} for a qualitative description of this spacetime. 

For $r_0>0$, \eqref{dsd} -- \eqref{kpd} describes a black string in ${\cal M}_3$. The winding and momentum of the string on the $\mathbb{S}^1$ are $p$ and $n$, respectively.  
Its energy above the ground state, \eqref{E}, is
\be\label{Ed}
E={Rvr_0^2\over 2\alpha'^2}\left(e^{-2\alpha_1}+\cosh 2\alpha_n\right)~,
\ee
in the limit described above. The entropy \eqref{S} is 
\be\label{Sd}
S={2\pi Rvr_5r_0^2\over\alpha'^2}\cosh\alpha_1\cosh\alpha_n~.
\ee
As we vary $r_0$, the properties of the black string change continuously between those of a BTZ black hole in AdS$_3$ for $r_0\ll r_1$ (i.e. for $E\ll{pR\over\alpha'}$)
to those of a $1+1$ dimensional $SL(2,\mathbb{R})/U(1)$ black hole in a linear dilaton spacetime for $r_0\gg r_1$ (i.e. for $E\gg{pR\over\alpha'}$), which becomes a black string in $2+1$ dimensions after we add back the $x$ circle. 

One can show that for general $E$ \eqref{Ed}, the entropy \eqref{Sd} takes the form 
\be\label{Sform}
S=\pi r_5\left(\sqrt{\left(E+{pR\over\alpha'}\right)^2-q_L^2}+\sqrt{\left(E+{pR\over\alpha'}\right)^2-q_R^2}\right)~, 
\ee
where 
\be\label{qlr}
(q_L,q_R)\equiv\left({n\over R}-{pR\over\alpha'},{n\over R}+{pR\over\alpha'}\right)~,
\ee
with $p$ and $n$ in \eqref{kpd}. For $E\ll pR/\alpha'$, the entropy \eqref{Sform} reduces to the Cardy entropy of a CFT with central charge $c_{\rm st}=6kp$,  $n=\bar L_0-L_0$ and $ER=L_0+\bar L_0-c_{\rm st}/12$. For $E\gg pR/\alpha'$, it gives a Hagedorn entropy $S=\beta_hE$, with 
\be\label{betah}
\beta_h= 2\pi r_5=2\pi\sqrt k l_s~.
\ee
As one varies the energy between these two extremes, it continuously interpolates between the two behaviors.  

In analyzing the background \eqref{dsd} -- \eqref{kpd}, it is sometimes useful to change coordinates from  $(t,r,x)$ to $(\tau,\rho,\varphi)$:
\be\label{newcoor}
\tau\equiv{r_5t\over R}~,\qquad\rho^2\equiv{R^2\over r_1^2-r_n^2}\left(r^2+r_n^2\right)~,\qquad\varphi\equiv{x\over R}~.
\ee
The spatial boundary coordinate $\varphi$ lives on a circle of unit radius,
\be\label{varphi}
\varphi\sim\varphi+2\pi~,
\ee
and the metric in the $(\tau,\rho,\varphi)$ directions and dilaton take the form
\be\label{metric}
ds^2=-{N^2\over 1+{\rho^2\over R^2}}d\tau^2+{d\rho^2\over N^2}+{\rho^2\over 1+{\rho^2\over R^2}}(d\varphi-N_\varphi d\tau)^2~,
\ee
and
\be\label{dilaton}
e^{2\Phi}={kv\over p}\sqrt{\left(1+{\rho_-^2\over R^2}\right)\left(1+{\rho_+^2\over R^2}\right)}{1\over 1+{\rho^2\over R^2}}~,
\ee
respectively. Here,
\be\label{NN}
N^2\equiv{(\rho^2-\rho_-^2)(\rho^2-\rho_+^2)\over r_5^2\rho^2}~,\qquad N_\varphi\equiv{\rho_-\rho_+\over r_5\rho^2}~,
\ee
and, \eqref{newcoor},
\be\label{rhopm}
\rho_+^2={r_0^2+r_n^2\over r_1^2-r_n^2}R^2~,\qquad\rho_-^2={r_n^2\over r_1^2-r_n^2}R^2.
\ee
The background \eqref{metric} -- \eqref{rhopm} describes a black string in $2+1$ dimensions, with an outer horizon at $\rho=\rho_+$ and an inner horizon at $\rho=\rho_-$, 
which correspond via \eqref{newcoor} to $r=r_0$ and $r=0$ in \eqref{dsd} -- \eqref{kpd}, respectively. Upon reduction on the $\varphi$ circle, it gives rise to a two-dimensional black hole with fundamental string winding and momentum $p$ and $n$, \cite{Giveon:2005mi}.

The inverse Hawking temperature, $\beta$, of the black string \eqref{metric} -- \eqref{rhopm} is
\be\label{betabh}
\beta={2\pi r_5\over r_0^2}\sqrt{(r_0^2+r_1^2)(r_0^2+r_n^2)}=2\pi R\sqrt{1+{\rho_+^2\over R^2}}{r_5\rho_+\over\rho_+^2-\rho_-^2}=\frac{2\pi r_5\tilde\rho_+\sqrt{(R^2+\tilde\rho_-^2)(R^2+\tilde\rho_+^2)}}{R(\tilde\rho_+^2-\tilde\rho_-^2)}~,
\ee
with
\be\label{tilderho}
\tilde\rho_\pm\equiv{\rho_\pm\over\sqrt{1+{\rho_\mp^2\over R^2}}}~.
\ee
Its Bekenstein-Hawking entropy, $S$, is~\footnote{The BH entropy and  Hawking temperature can be obtained from \eqref{S} and \eqref{betaa} by taking the near-$NS5$ limit, or directly from \eqref{metric}, \eqref{NN}, using standard techniques.}
\be\label{Sbh}
S={2\pi Rvr_5\over\alpha'^2}\sqrt{(r_0^2+r_1^2)(r_0^2+r_n^2)}={2\pi kp\over r_5}{\rho_+\over\sqrt{1+{\rho_-^2\over R^2}}}={2\pi kp\over r_5}\tilde\rho_+~.
\ee
The momentum $n$ in \eqref{kpd} is given in terms of $\rho_\pm$ by
\be\label{nbh}
n={p\over\alpha'}{\rho_-\rho_+\over\sqrt{\left(1+{\rho_-^2\over R^2}\right)\left(1+{\rho_+^2\over R^2}\right)}}\equiv{p\over\alpha'}\tilde\rho_-\tilde\rho_+~,
\ee
where $\tilde\rho_\pm$ are defined in \eqref{tilderho}, and $p$ is the winding charge in \eqref{kpd}. The energy of the black string, $E$, can be obtained from the ADM construction in the near-$NS5$ limit, \eqref{Ed},
or from its temperature, $T\equiv 1/\beta$,  \eqref{betabh}, and entropy, \eqref{Sbh}, via the thermodynamic relation
\be\label{Evia}
E=\int_0^S T(S')dS'~,
\ee
with $n$ fixed.
The integration constant is determined by the condition that $E=0$ when $\rho_+=0$. The result of the integral in \eqref{Evia} is 
\be\label{Ebh}
E={pR\over\alpha'}\left(-1+\sqrt{\left(1+{\tilde\rho_-^2\over R^2}\right)\left(1+{\tilde\rho_+^2\over R^2}\right)}\right)~.
\ee
So far we discussed the near-horizon geometry of $k$ $NS5$-branes in the presence of $p$ fundamental strings. In the next subsection, we briefly review its boundary interpretation.

\subsection{The boundary theory}
\label{twothree}

One can study the holographic dual of string theory on ${\cal M}_3\times \mathbb{S}^3\times \mathbb{T}^4$ from two different points of view. From the UV perspective, ${\cal M}_3$ is a deformation of its large $r$ (or, equivalently, large $\rho$) limit, which is a linear dilaton spacetime. Its boundary dual is $5+1$ dimensional LST (see e.g. \cite{Aharony:1999ks,Kutasov:2001uf} for reviews), compactified down to $1+1$ dimensions on a $\mathbb{T}^4$, in a state with a large number, $p$, of fundamental strings. From the IR point of view, ${\cal M}_3$ is a (UV) deformation of its small $r$ (or $\rho$) limit, which is AdS$_3$. Thus, it is holographically dual to an irrelevant deformation of the CFT$_2$ dual of string theory on AdS$_3\times \mathbb{S}^3\times \mathbb{T}^4$. 

Superficially, it might seem that the second perspective on the theory is less useful, since irrelevant deformations are usually not under control, however, it turns out that this is not the case here, for two reasons. One is that LST, the non-local theory that underlies the first point of view, is still not well understood. In fact, one of the reasons for the interest in single trace $T\bar T$ is to gain additional information on LST. The second is that the particular irrelevant deformation that enters the second point of view is better behaved than generic irrelevant deformations of a CFT. In fact, the main gap in the understanding of this deformation is our limited understanding of the undeformed CFT$_2$. We next comment on a few features of this second description of the theory, referring the reader to the literature for a more detailed discussion. 

As mentioned earlier, a lot is known about the CFT$_2$ dual to string theory on AdS$_3\times \mathbb{S}^3\times \mathbb{T}^4$. In particular, it is known that the perturbative string states in the bulk theory are described on the boundary by a symmetric product theory. Denoting by $M_{6k}$ the CFT\footnote{Here $\mathbb{R}_\phi$ is a linear dilaton CFT, with slope determined by the requirement that the total central charge of $M_{6k}$ is 6k. The $\mathbb{S}^3$ factor is described by a superconformal $SU(2)$ WZW model, with (total) level $k$.} $M_{6k}=\mathbb{R}_\phi\times \mathbb{S}^3\times \mathbb{T}^4$, it is known that the CFT $\left(M_{6k}\right)^N/S_N$, where $N$ is any sufficiently large integer, describes all the $\delta$-function normalizable perturbative string states  \cite{Argurio:2000tb,Giveon:2005mi,Chakraborty:2025nlb}. To describe the normalizable states, one needs to deform the above symmetric product theory by a $\mathbb{Z}_2$ twisted modulus, which creates a wall in the linear dilaton direction $\mathbb{R}_\phi$  \cite{Balthazar:2021xeh,Eberhardt:2021vsx,Chakraborty:2025nlb}. The aforementioned normalizable states live in the vicinity of the wall.

The deformation from AdS$_3$ to ${\cal M}_3$, reviewed in the previous subsections, corresponds in the symmetric product $(M_{6k})^N/S_N$ to adding to the Lagrangian of the seed CFT $M_{6k}$ the $T\bar T$ deformation \cite{Giveon:2017nie,Giveon:2017myj}. In the full symmetric product, it thus corresponds to adding to the Lagrangian the deformation $t\sum_{i=1}^N T_i\bar T_i$. This form is the origin of the name {\it single trace $T\bar T$ deformation}. The usual $T\bar T$ deformation is $t\sum_{i,j=1}^N T_i\bar T_j=tT\bar T$, and is referred to in the symmetric product as {\it double trace $T\bar T$ deformation}, since it is a product of two single trace operators, $T=\sum_i T_i$ and $\bar T=\sum_i \bar T_i$. Another reason for the names is that in string theory on AdS$_3$, the single trace deformation is local on the worldsheet -- it is described by adding to the worldsheet Lagrangian a particular vertex operator constructed in \cite{Kutasov:1999xu}. On the other hand, the double trace deformation is non-local on the worldsheet, and corresponds to adding to the worldsheet action a product of two operators, separately integrated over the worldsheet. 

It is important to stress that while the deformed symmetric product description is valid for perturbative string states, it is known to fail, already in the AdS$_3$ background before the deformation, for states with energy much larger than the spacetime central charge $c_{\rm st}=6kp$. Such states in the CFT correspond  in the bulk to BTZ black holes, and the question what are the states in the boundary CFT that describe them is essentially the question what are the microstates of these black holes. This question remains unresolved as of this writing. 

Thus, when studying high energy states, one can only appeal to what one can surmise from  thermodynamics. In particular, the Cardy entropy of the boundary theory agrees with the Bekenstein-Hawking entropy of BTZ black holes \cite{Strominger:1997eq}, and one can generalize that discussion from AdS$_3$ to ${\cal M}_3$. This leads \cite{Giveon:2017nie, Chakraborty:2023mzc,Chakraborty:2023zdd} to a formula for the deformed energies of high energy states in terms of the energies of the undeformed states and the coupling. We can define a dimensionless coupling, 
\be\label{deflambda}
\lambda_d\equiv{\alpha'\over pR^2}={4r_5G_3\over R^2}~,
\ee
where $R$ is the radius of the spatial circle \eqref{xR} and $\alpha'/p$ plays the role of the $T\bar T$ coupling. The second equality rewrites it in a form that will be useful later, in terms of the three dimensional Newton constant, $G_3=r_5/4kp$, and the AdS radius $r_5$. Plugging \eqref{deflambda} into \eqref{Ebh}, one finds \cite{Giveon:2017nie,Chakraborty:2023mzc,Chakraborty:2023zdd} that the deformed energy is given by 
\be\label{Elambda}
RE(\lambda_d)={1\over\lambda_d }\left(-1+\sqrt{1+2\lambda_d E(0)R+\left(\lambda_d RP\right)^2}\right)~,
\ee
where
\be\label{PnR}
P\equiv{n\over R}~,
\ee
and $E(0)$ is the energy of the state before turning on the deformation.

As discussed in \cite{Giveon:2017nie,Chakraborty:2023mzc,Chakraborty:2023zdd}, single trace $T\bar T$ deformed CFT has the property that the entropy at a given $\lambda_d$ \eqref{deflambda}, and energy $E(\lambda_d)$ \eqref{Elambda}, is independent of $\lambda_d$, i.e.
\be\label{Sfixed}
S(\lambda_d, E(\lambda_d),P)=S(0, E(0),P)=2\pi\sqrt{c_{\rm st}\over 12}\left(\sqrt{(E(0)+P)R}+\sqrt{(E(0)-P)R}\right)
\ee
for any $\lambda_d$. This implies that the entropy \eqref{Sfixed} interpolates between Cardy and Hagedorn behavior as the energy increases. Indeed, for small $\lambda_d E(0) R$, the energy \eqref{Elambda} satisfies $E(\lambda_d)\simeq E(0)$, while for large $\lambda_d E(0) R$, $E(\lambda_d)\sim \sqrt{E(0)}$. Since the (Cardy) entropy in the undeformed theory goes like $S\sim\sqrt{E(0)}$, \eqref{Sfixed}, we see that in the deformed theory it behaves as indicated above; see also the discussion around \eqref{Sform}. 

\section{Negative double-trace $T\bar T$}
\label{secthree}

In the previous section, we described the single trace $T\bar T$ deformation, a particular UV deformation of string theory on AdS$_3$ that has much in common with the $T\bar T$ deformation of the dual CFT$_2$, but is distinct from it. It is natural to ask how the actual $T\bar T$ deformation, which we  referred to as  double trace $T\bar T$, arises in the bulk theory. There has been a lot of work on this question, starting with \cite{McGough:2016lol}. In this section, we will briefly review their construction. As mentioned above, it and the results reviewed in section \ref{sectwo} will serve as the starting point of this work. 

The basic claim of \cite{McGough:2016lol} is that turning on the double trace $T\bar T$ deformation with negative coupling corresponds in the bulk to introducing a finite radial cutoff in AdS$_3$, at some  $\rho=\rho_b$ in the background \eqref{metric} -- \eqref{NN}. There has been a lot of discussion of this procedure in the literature (see e.g. \cite{Kraus:2018xrn,Guica:2019nzm}). The conclusion seems to be that it is valid in the gravity approximation, but when one adds  matter fields, in general new issues arise regarding the boundary conditions for these fields. This issue will come up in our discussion below. The cutoff description is useful for describing the physics of black holes, but for generic states, like perturbative string states, one must go beyond it.

\subsection{AdS$_3$/CFT$_2$}

From the point of view of the discussion in section \ref{sectwo}, BTZ black holes in AdS$_3$ are described by geometries in which the horizon is deep in the AdS$_3$ region in ${\cal M}_3$. In the coordinates \eqref{dsd}, this means that $r, r_0, r_n\ll r_1$. In  \eqref{metric}, \eqref{dilaton} it means that $\rho,\rho_\pm\ll R$. In this limit, these equations take the form
\be\label{btzmetric}
ds^2=-N^2d\tau^2+{d\rho^2\over N^2}+\rho^2(d\varphi-N_\varphi d\tau)^2~,
\ee
and 
\be\label{btzdilaton}
e^{2\Phi}={kv\over p}~.
\ee
Here,
\be\label{btzNN}
\begin{split}
&N^2\equiv{(\rho^2-\rho_-^2)(\rho^2-\rho_+^2)\over r_5^2\rho^2}={\rho^2\over r_5^2}-8G_3M+{(4G_3n)^2\over\rho^2}~,\\ &N_\varphi\equiv{\rho_-\rho_+\over r_5\rho^2}={4G_3n\over\rho^2}~,\qquad G_3\equiv {r_5\over 4kp}~.
\end{split}
\ee
The background \eqref{btzmetric} -- \eqref{btzNN} describes a BTZ black hole of mass $M$ and angular momentum $n$ in an AdS$_3$ spacetime with $R_{\rm AdS}=r_5$.
The horizon positions $\rho_\pm$ are given in the limit by
\be\label{rhopmG}
\rho_\pm^2=4r_5G_3\left(r_5M\pm\sqrt{(r_5M)^2-n^2}\right)~,
\ee
or, equivalently, 
\be\label{MJ}
M=\frac{\rho_+^2+\rho_-^2}{8r_5^2G_3}~,\qquad n=\frac{\rho_+\rho_-}{4r_5G_3}~.
\ee
The Bekenstein-Hawking entropy of the black hole is
\be\label{btzS}
S={\pi\rho_+\over 2G_3}={2\pi kp\over r_5}\rho_+~.
\ee
From the point of view of section \ref{sectwo}, \eqref{btzS} corresponds to the small $\rho/R$ limit of \eqref{Sbh}. It can be written in the form
\be\label{Scft}
S=\pi\sqrt{{r_5\over G_3}\left(r_5M+\sqrt{(r_5M)^2-n^2}\right)}=2\pi\left(\sqrt{{c_{\rm st}\over 12}\left(r_5M-n\right)}+\sqrt{{c_{\rm st}\over 12}\left(r_5M+n\right)}\right)~,
\ee
where
\be\label{ccft}
c_{\rm st}={3r_5\over 2G_3}=6kp~
\ee
is the central charge of the spacetime or boundary CFT.  The standard AdS/CFT map,
\be\label{L0}
L_0+\bar L_0-{c_{\rm st}\over 12}= r_5M~,\qquad\bar L_0-L_0= n~,
\ee
relates \eqref{Scft} to the Cardy entropy of the boundary CFT,
\be\label{Scardy}
S=2\pi\sqrt{c_{\rm st}\over 6}\left(\sqrt{L_0-{c_{\rm st}\over 24}}+\sqrt{\bar L_0-{c_{\rm st}\over 24}}\right)~.
\ee

Introducing a radial UV cutoff at $\rho=\rho_b$ can be thought of as ``moving the CFT into the bulk'' (to quote \cite{McGough:2016lol}). In other words, the theory dual to gravity in the cavity $\rho\le\rho_b$ lives at the wall $\rho=\rho_b$. Thus, one can identify the physical parameters of the boundary theory from the geometry \eqref{btzmetric}. In \cite{McGough:2016lol}, this was done, following \cite{Brown:1994gs}, in the microcanonical ensemble. In this section, we will rederive their results in the canonical ensemble, following \cite{York:1986it}. This will be useful in the next section.

We start by Wick rotating \eqref{btzmetric} to Euclidean time, $\tau_E=i\tau$. As is standard, $\tau_E$ must be periodic for regularity of the Euclidean geometry at $\rho=\rho_+$. For zero angular momentum, its period is $\beta_{\rm BTZ}=2\pi r_5^2/\rho_+$. For general angular momentum, to get a real metric one needs to Wick rotate the  horizon angular velocity $\Omega_H=\rho_-/(r_5\rho_+)$ as well, and define 
\be\label{defome}
\Omega_E=-i\Omega_H=\frac{\rho^E_-}{r_5\rho_+}~,
\ee
where we formally Wick rotated $\rho_-$, and defined $\rho_-^E=-i\rho_-$. 

Regularity of the Euclidean geometry requires in this case the combined identification 
\be\label{combid}
(\tau_E,\varphi)\sim (\tau_E+\beta_{\rm BTZ}, \varphi+\beta_{\rm BTZ}\Omega_E)~,
\ee
where $\beta_{\rm BTZ}$ is the inverse temperature of the BTZ black hole,
\be\label{betabtz}
\beta_{\rm BTZ}=\frac{2\pi r_5^2\rho_+}{\rho_+^2+(\rho_-^E)^2}~.
\ee
This, together with the identification \eqref{varphi}, implies that the space parametrized by $(\tau,\varphi)$ is a torus. The form of the metric \eqref{btzmetric} near the boundary at large $\rho$ makes it natural to define the complex coordinate 
\be\label{defz}
z=\varphi+i\tau_E/r_5~,
\ee
in terms of which the  metric at a fixed large $\rho$ is $ds^2=\rho^2dzd\bar z$, and the identifications (\ref{varphi}), \eqref{combid} are $z\sim z+2\pi$, $z\sim z+2\pi\tau_{\rm st}$, with 
\be\label{taust}
2\pi\tau_{\rm st}=\beta_{\rm BTZ}\Omega_E+i\beta_{\rm BTZ}/r_5~.
\ee

The Euclidean BTZ black hole contributes to the partition sum in the boundary CFT
\be\label{partsum}
Z(\hat\beta, \hat \mu)={\rm Tr} q^{L_0-{c_{\rm st}\over24}}{\bar q}^{\bar L_0-{c_{\rm st}\over24}},\;\;\;
\ee
where
\be\label{defqmu}
 q=e^{2\pi i\tau_{\rm st}} =e^{-\hat \beta(1+i\hat \mu)} \ ,
\ee
with $\hat\beta=\beta_{\rm BTZ}/r_5$ and $\hat \mu=-r_5\Omega_E$. If we fix the modulus of the boundary torus, $\tau_{\rm st}$, we can use \eqref{defome}, \eqref{betabtz}, \eqref{taust} to determine the parameters of the corresponding bulk black hole, $\rho_+$ and $\rho_-^E$. Recall that the  dimensionless quantity $\hat\beta$ defined in \eqref{defqmu} corresponds in the boundary CFT to the inverse temperature measured in units of the radius of the spatial circle. On the other hand, it follows from (\ref{L0}) that the inverse temperature conjugate to the BTZ mass $M$ is $\beta_{\rm BTZ}$.

\subsection{Radial cavity in AdS$_3$ and its dual}\label{sec:BTZb}

So far, we discussed the situation in the theory without a UV cutoff, i.e. we took $\rho_b=\infty$. We now put back the cutoff, to connect to the discussion of \cite{McGough:2016lol}. The boundary metric is now evaluated at $\rho=\rho_b$, 
\be\label{metrob}
ds^2_{\rho_b}=N_E^2(\rho_b)d\tau_E^2+\rho_b^2\left(d\varphi-N^E_\varphi(\rho_b) d\tau_E\right)^2~,
\ee
where 
\be\label{eucNN}
\left(N_E(\rho_b)\right)^2={(\rho_b^2+(\rho_-^E)^2)(\rho_b^2-\rho_+^2)\over r_5^2\rho_b^2},\;\;\;N^E_\varphi(\rho_b)={\rho_-^E\rho_+\over r_5\rho_b^2}~.
\ee
We can write \eqref{metrob} in the form 
\be\label{boundmet}
ds^2_{\rho_b}=\rho_b^2d\tilde zd\bar{\tilde z}~,
\ee
where 
\be\label{defzzz}
\tilde z=\varphi-N^E_\varphi(\rho_b)\tau_E+i\frac{N_E(\rho_b)}{\rho_b}\tau_E~.
\ee
The two identifications \eqref{varphi}, \eqref{combid}, correspond in terms of $\tilde z$ to $\tilde z\sim \tilde z+2\pi$ and $\tilde z\sim \tilde z+2\pi\tilde \tau_{\rm st}$, respectively, where
\be\label{tautilde}
\tilde\tau_{\rm st}=\frac{\beta_{\rm BTZ}}{2\pi}\left(\Omega_E-N^E_\varphi(\rho_b)+i\frac{N_E(\rho_b)}{\rho_b}\right)~.
\ee
We note in passing that in the limit $\rho_b\to\infty$, the modulus \eqref{tautilde} simplifies. From eq. \eqref{eucNN} we see that in this limit $N^E_\varphi(\rho_b)\to 0$, and $N_E(\rho_b)/\rho_b\to 1/ r_5$. Thus, in the limit we recover the modulus of the torus in the original, unperturbed CFT \eqref{taust}. 

Another interesting limit is $\rho_b\to\rho_+$. In this limit, the twist of the two-torus (the real part of $\tilde{\tau}_{\rm st}$) vanishes, and $N_E(\rho_b)\to 0$. Since the real part of $\tilde{\tau}_{\rm st}$ goes to zero faster than the imaginary part, in this limit $\tilde{\tau}_{\rm st}\to 0$ along the imaginary axis. Thus, we see that the torus untwists and pinches off. The radius of the thermal circle goes to zero, while that of the spatial circle, $\rho_+$, remains finite. 

From the torus geometry \eqref{boundmet}, \eqref{tautilde}, we can read off the physical parameters of the boundary theory. The first identification of $\tilde z$ listed above corresponds to \eqref{varphi}. From it and \eqref{boundmet}, we can read off the physical radius of the boundary spatial circle, $R_b$,
\be\label{Rb}
R_b=\rho_b~.
\ee
The inverse temperature of the boundary theory is given by $\beta(\rho_b)=2\pi\rho_b{\rm Im}{\Tilde\tau}_{\rm st}$, which corresponds to the temperature 
\be\label{Tbtz}
\beta(\rho_b)={2\pi r_5\rho_+\sqrt{(\rho_b^2+\left(\rho^E_-\right)^2)(\rho_b^2-\rho_+^2)}\over(\rho_+^2+\left(\rho^E_-\right)^2)\rho_b }~.
\ee
The real part of $\tilde{\tau}_{\rm st}$ \eqref{tautilde} determines the chemical potential that couples to angular momentum, \eqref{partsum}, via the relation $\mu(\rho_b)=-{\rm Re}\tilde\tau_{\rm st}/(R_b{\rm Im}\tilde\tau_{\rm st}) $, yielding
\be\label{chemmu}
\mu(\rho_b)=-\frac{\rho_-^E}{R_b\rho_+}\sqrt{\frac{\rho_b^2-\rho_+^2}{\rho_b^2+(\rho_-^E)^2}}\ .
\ee
The way to think about equations \eqref{Rb} -- \eqref{chemmu}, is the following. To study the boundary theory at a particular temperature $\beta$ and chemical potential $\mu$, on a spatial circle of a particular radius $R_b$, we set $\rho_b$ to the value \eqref{Rb}, and solve \eqref{Tbtz}, \eqref{chemmu} for $\rho_\pm$ with $\beta(\rho_b)=\beta$ and $\mu(\rho_b)=\mu$. The resulting $\rho_\pm$ can then be used to calculate the mass $M$ and (imaginary) angular momentum $n$ of the Euclidean BTZ black hole that contributes to the partition sum at that temperature and chemical potential, using \eqref{rhopmG}. 

To implement this procedure, we start with the case of vanishing chemical potential, $\mu=0$, i.e. $\rho_-^E=0$, \eqref{chemmu}. Plugging this into (\ref{Tbtz}), and eliminating  $\rho_+$ in favor of the entropy $S$, using \eqref{btzS}, we find the relation 
\be\label{TofS0}
\beta(S)=\frac{\pi r_5\sqrt{(\pi R_b)^2-(2G_3S)^2}}{G_3S}
\ee
for the inverse temperature as a function of the entropy. From this relation one can calculate the energy conjugate to $\beta$ in the usual way. The Euclidean gravity path integral for the partition sum (\ref{partsum}) computes 
\begin{equation}
\label{partsumintE}
    \int dEe^{S(E)-\beta E}\ .
\end{equation}
In the semiclassical approximation, the integral over energy is dominated by a saddle point, at which  
\begin{equation}
\label{sdpt}
    \frac{\partial S}{\partial E}=\beta\ .
\end{equation}
Combining \eqref{TofS0} and \eqref{sdpt} we find 
\begin{equation}
    \frac{\partial S}{\partial E}=\frac{\pi r_5\sqrt{(\pi R_b)^2-(2G_3S)^2}}{G_3S}\ ,
\end{equation}
a differential equation for the energy, whose solution (with the initial condition $E(0)=0$) is
\begin{equation}
\label{EoS}
    E=\frac{R_b}{4 r_5G_3}\left(1-\sqrt{1-\left(\frac{2G_3S}{\pi R_b}\right)^2}\right)\ .
\end{equation}
As we will see below, the momentum $n$ at the saddle point vanishes for $\mu=0$. Thus, by using 
(\ref{Scft}), we can express the entropy in terms of $M$, and find 
\begin{equation}
\label{Ebtzn0}
    E=\frac{R_b}{4 r_5G_3}\left(1-\sqrt{1-\frac{8r_5^2G_3M}{R_b^2}}\right)\ .
\end{equation}

It is not hard to generalize the above discussion to  $\mu\not=0$.  (\ref{partsumintE}) takes in this case the form
\begin{equation}
    \int dE \int dn\, e^{S(E,n)-\beta (E+i\mu n)}\ .
\end{equation}
The saddle point thus satisfies the conditions
\begin{equation}
\label{dSdEdSdn}
    {\partial S\over \partial E}\Big|_n=\beta~,\qquad {\partial S\over \partial n}\Big|_E=i\beta\mu\ .
\end{equation}
To solve the two equations, we express the temperature (\ref{Tbtz}) and chemical potential (\ref{chemmu}) in terms of entropy and angular momentum, by using (\ref{MJ}) and \eqref{btzS}, 
\be\label{TofSn}
\beta(S,n)={ \pi r_5G_3S^2\sqrt{[(R_bS)^2-(2n\pi r_5)^2][(\pi R_b)^2-(2G_3S)^2]}\over R_b[(G_3S^2)^2-(n\pi^2r_5)^2]}~,
\ee
\be\label{muofSn}
\mu(S,n)={in\pi r_5\sqrt{(\pi R_b)^2-(2G_3S)^2}\over R_bSG_3\sqrt{(R_bS)^2-(2n\pi r_5)^2}}~.
\ee
Note that (\ref{muofSn}) shows that $n=0$ when $\mu=0$, as expected.

Integrating the first equation in (\ref{dSdEdSdn}), with (\ref{TofSn}), we obtain the energy
\be
\begin{split}
E=\frac{R_b}{4r_5G_3}\left[
C(n)-\sqrt{\left(1-\left(\frac{2n\pi r_5}{R_bS}\right)^2\right)
\left(1-\left(\frac{2G_3S}{\pi R_b}\right)^2\right)}
\right]~,
\end{split}
\ee
where $C(n)$ is an undetermined function of $n$ that is independent of $S$. To determine $C(n)$, we hold $E$ fixed and differentiate the above expression with respect to $n$. This gives
\begin{equation}\label{nmu}
\begin{split}
\left.\frac{\partial S}{\partial n}\right|_E
={}&
\frac{nS\pi^2r_5^2}{R_b^2}
\frac{(\pi R_b)^2-(2G_3S)^2}
{(n\pi^2r_5)^2-(G_3S^2)^2}
\\
&+
\frac{\pi S^2
\sqrt{(\pi R_b)^2-(2G_3S)^2}
\sqrt{(R_bS)^2-(2n\pi r_5)^2}}
{4\left(n^2\pi^4r_5^2-G_3^2S^4\right)}
C'(n)\ .
\end{split}
\end{equation}
Consistency with the second equation in (\ref{dSdEdSdn}), after substituting (\ref{TofSn}) and (\ref{muofSn}) into its r.h.s., requires $C'(n)=0$. Hence, $C(n)$ is a constant. We fix this constant by requiring that, when $\mu=0$, or equivalently $n=0$ as discussed above, the energy reduces to (\ref{Ebtzn0}). We  find
\be\label{Ebtz}
E={R_b\over4r_5G_3}
\left(
1-\sqrt{
1-{8r_5^2G_3M\over R_b^2}
+\left({4r_5G_3n\over R_b^2}\right)^2
}
\right)~.
\ee
One may also verify that, in the extremal case $\rho_+=\rho_-$, or equivalently $r_5M=n$, one has $E=n/R_b$.

To put \eqref{Ebtz} in a more suggestive form, we define the dimensionless $T\bar T$ coupling 
\be\label{lambdagravity}
\lambda_t\equiv -{4r_5G_3\over R_b^2}~, 
\ee
and use the fact that 
\be\label{E0Rb}
R_bE(0)=r_5M~,\;\;\; PR_b= n~.
\ee
Plugging this into \eqref{Ebtz}, we find 
\be\label{Ebndry}
R_b E(\lambda_t)={1\over\lambda_t}\left(-1+\sqrt{1+2\lambda_t E(0)R_b+(\lambda_t PR_b)^2}\right)~,
\ee
which is precisely the spectrum of a $T\bar T$ deformed CFT with the coupling $t=-4r_5G_3=-\alpha'/p$. In particular, when the coupling \eqref{lambdagravity} goes to zero, the energy \eqref{Ebndry} goes to the undeformed one, $\lim_{\lambda_t\to 0} E(\lambda_t)=E(0)$.  

An interesting feature of the above analysis concerns the entropy of  caged BTZ black holes. Since the entropy is a feature of the horizon, it is given by eq. \eqref{btzS} which only depends on the coupling $\lambda_t$ and the energy $E(\lambda_t)$ through the corresponding undeformed energy $E(0)$, via \eqref{rhopmG}, \eqref{E0Rb}. That means that if we change $\lambda_t$ \eqref{lambdagravity} and $E(\lambda_t)$ \eqref{Ebndry}, while keeping the size of the circle $R_b$ and the undeformed energy $E(0)$ fixed, the entropy does not change. In other words, the entropy satisfies the property \eqref{Sfixed} noted above for black holes in ${\cal M}_3$. This is consistent with the picture presented in \cite{Aharony:2018bad}, where it was noted that (the first equality in) eq. \eqref{Sfixed} is valid in any theory in which the change of the energies of states as a function of a coupling only depends on the energy of the states, and not on any of their other attributes. Introducing the cavity for black holes is clearly such a deformation (essentially due to the no-hair theorem), so it's not surprising that it satisfies \eqref{Sfixed}.  

It is also useful to note for future reference that the bulk radial cutoff at $\rho=\rho_b$ implies an upper bound on the energies of the states. From the bulk point of view, this bound comes from the requirement that the black hole fits in the cavity, i.e. $\rho_+<\rho_b$. In the boundary theory, it corresponds to the fact that the energy \eqref{Ebtz} satisfies the bound 
\begin{equation}
\label{Emaxa}
    E<E_{\rm max}\equiv\frac{\rho_b}{4r_5G_3}={1\over |\lambda_t|R_b}~.
\end{equation}
There is also a maximal entropy,
\be\label{Smaxa}
S<S_{\rm max}={\pi\rho_b\over 2G_3}~,
\ee
that corresponds to the maximal energy, $S_{\rm max}=S(E_{\rm max})$.

\section{Radial cavity in ${\cal M}_3$ }
\label{secfour}

In the previous two sections, we described two different, but related, deformations of AdS$_3$, one corresponding in the bulk to replacing AdS$_3$ by ${\cal M}_3$ (section \ref{sectwo}), the other to introducing a radial cutoff in AdS$_3$ (section \ref{secthree}). We explained the interpretation of these deformations in the boundary theory in terms of certain irrelevant deformations of the CFT$_2$. 

The question we would like to address in this section is what happens when we combine the two. In other words, we will study the bulk and boundary theories that arise when we introduce a UV cutoff $\rho\le \rho_b$ in \eqref{metric} -- \eqref{NN} without taking the limit $\rho\ll R$. The theories described in sections \ref{sectwo}, \ref{secthree} correspond to different limits of this theory. 

One's first thought might be that the interpolating theory corresponds to turning on a combination of single and double trace $T\bar T$ deformations with general non-zero couplings \eqref{deflambda}, \eqref{lambdagravity}, respectively. We will show that while this is one possibility, the actual space of deformations is much richer and more interesting. We will do this by presenting a large set of possible boundary conditions, and computing the spectrum of black holes in some examples. We will focus on states with vanishing spatial momentum, $n=0$, for which the analysis simplifies significantly. The generalization to arbitrary $n$ will be left for future work.

The starting point of our discussion is the geometry describing a black hole in ${\cal M}_3$, \eqref{metric} -- \eqref{NN}, with the constraint $\rho\le \rho_b$.  We will set the chemical potential $\mu$ to zero, which leads, as in section \ref{secthree}, to vanishing momentum, $n=0$; thus, we will take $N_\varphi=\rho_-=0$. We will continue denoting the (single) horizon of the black hole by $\rho_+$. 

The background \eqref{metric} -- \eqref{NN} depends in this case on three parameters (in addition to the numbers of branes, $(k, p$), and the string scale): $(\rho_b, R,\rho_+)$. These parameters can be mapped to the three non-trivial parameters of the boundary theory -- the size of the spatial circle on which the theory is defined, $R_b$, the inverse temperature $\beta$, and the value of the dilaton at the boundary, $\Phi_b=\Phi(\rho_b)$. 

Looking back at \eqref{metric} -- \eqref{NN}, we have 
\be\label{Rbb}
R_b=R\frac{\rho_b}{\sqrt{\rho_b^2+R^2}}~,
\ee
\be\label{TM3}
\beta={2\pi r_5\over\rho_+}\sqrt{\rho_b^2-\rho_+^2}{\sqrt{\rho_+^2+R^2}\over\sqrt{\rho_b^2+R^2} }~,
\ee
\begin{equation}
\label{Phib}
    \chi\equiv {p\over kv}e^{2\Phi_b}=
{\sqrt{\rho_+^2+R^2}\over\sqrt{\rho_b^2+R^2} }   
{1\over \sqrt{1+{\rho_b^2\over R^2}}}\ .
\end{equation}
Note that the physical range of $\chi$ is $0\le\chi\le1$, in agreement with its definition, \eqref{dilaton} evaluated at $\rho=\rho_b$. For a cavity deep inside the AdS$_3$ region, $\chi$ takes the value $\chi\simeq 1$, its upper bound. Conversely, as the boundary $\rho_b\to\infty$, $\chi$ approaches its lower bound $(\chi=0)$. These two limits correspond to the discussions in sections \ref{secthree} and \ref{sectwo}, respectively.  Note also that eq. \eqref{Rbb} implies that $\rho_b>R_b$ for all finite $R$. Equality is achieved when $R\to\infty$, i.e. when the cavity is located deep inside the AdS$_3$ region. This observation will be useful later.

One can think about the three equations \eqref{Rbb} -- \eqref{Phib}
in the following way. The first says that a particular combination of the two parameters $(\rho_b,R)$, namely, the one on the r.h.s. of \eqref{Rbb}, is a feature of the boundary theory (the radius of the circle on which that theory is defined), and not of the particular state in that theory that we are considering. This might seem obvious, but  we will see that in general $\rho_b$ and $R$ separately do depend on the state. Equation \eqref{TM3} states that a particular combination of the three parameters $(\rho_b, R,\rho_+)$ gives the temperature in the boundary theory, which of course does depend on the state. 

Equation \eqref{Phib}  determines the boundary value of the dilaton $\Phi$ in terms of the above three bulk parameters, but by itself it does not provide a constraint on those parameters. To specify the boundary theory, one needs to provide the boundary conditions for $\Phi$,  independently of \eqref{Phib}. One possible boundary condition is Dirichlet: $\Phi_b$, or equivalently $\chi$ \eqref{Phib}, fixed, and in particular independent of the state. We will start by discussing the theory obtained by imposing this boundary condition. Later in the section, we will discuss a large class of additional possible boundary conditions and their consequences. 

Before continuing to that discussion, we would like to note that the freedom of imposing different boundary conditions, that we will utilize in this section, is absent in the limit where the system reduces to that of section \ref{secthree}. As discussed above, this limit corresponds to taking $\rho_b\ll R$. In that limit, equations \eqref{Rbb} and \eqref{Phib} simplify. The former takes the form \eqref{Rb}, which means that $\rho_b$ is fixed (i.e. independent of the state) and equal to the spatial radius in the boundary theory. The latter takes the form $\chi=1$, which means that the dilaton is fixed as well. Of course, in this case, it is not only the dilaton at the boundary that is fixed -- the dilaton takes the fixed value \eqref{btzdilaton} throughout the cavity. 

The quantity $R$ \eqref{xR} decouples in this limit, and one can think of equations \eqref{Rbb} and \eqref{TM3} as determining $\rho_b$ and $\rho_+$ in terms of $R_b$ and $\beta$, with equation \eqref{Phib} being redundant. 
Equations \eqref{Rbb} and \eqref{TM3} take the form \eqref{Rb} and \eqref{Tbtz} (with $\rho_-^E=0$), respectively, and the analysis below reduces to that of section \ref{secthree}. Thus, the freedom we will discuss below owes its existence to the generalization from AdS$_3$ to ${\cal M}_3$ or, directly in ${\cal M}_3$, from $\rho_b$ which is deep in the AdS$_3$ region to general $\rho_b$. 

\subsection{Dirichlet boundary conditions}
\label{Dirichlet}

\begin{figure}
	\centering
\includegraphics[scale=0.7]{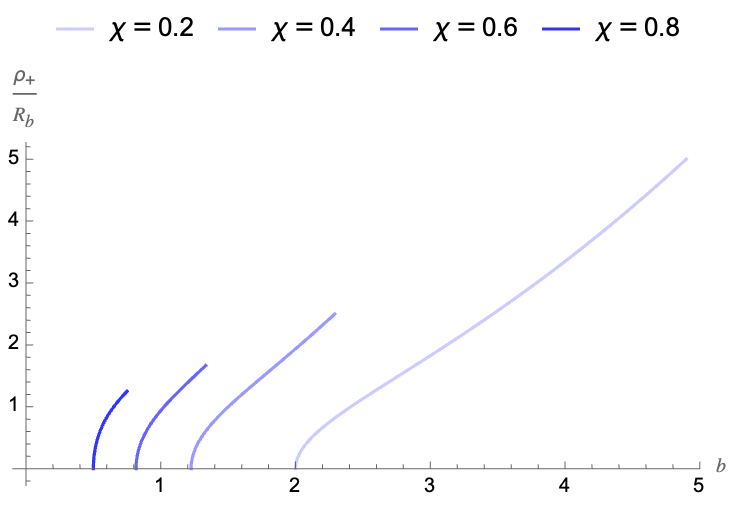}
\caption{\label{rhopbfig} $\rho_+$ as a function of $b$ with fixed $\chi$, (\ref{chib}).}
\end{figure}

Fixing the dilaton on the boundary of the spherical cavity in ${\cal M}_3$, eq. \eqref{Phib} implies that $(\rho_b,R)$ depend on $\rho_+$, such that the combinations on the r.h.s. of \eqref{Rbb}, \eqref{Phib} are independent of this variable. It is convenient to describe that dependence in terms of the dimensionless parameter $b(\rho_+)$, 
\be\label{defrhobb}
b=\frac{\rho_b}{R}~.
\ee
Plugging \eqref{defrhobb} into \eqref{Phib}, we find that 
\begin{equation}
\label{chib}
    \frac{\rho_+}{R_b}=\frac{\sqrt{1+b^2}}{b}\sqrt{(1+b^2)^2\chi^2-1}\ .
\end{equation}
A few comments about these equations:
\begin{itemize}
\item The way to read eq. \eqref{chib} is as determining $b(\rho_+)$, for fixed $R_b$ \eqref{Rbb} and $\chi$ \eqref{Phib}.
\item By its definition \eqref{defrhobb}, $b$ runs from $0$ to $\infty$, but according to \eqref{chib}, for fixed $\chi$ it is bounded from below,
\begin{equation}\label{lowb}
b\ge b_{\rm min}=\sqrt{\frac{1}{\chi}-1}\ . \end{equation}
Recall that $\chi\le1$, so the r.h.s. of \eqref{lowb} is real. The value $b_{\rm min}$ corresponds to $\rho_+=0$.
\item There is also an upper bound on $b$, that corresponds to $\rho_+=\rho_b$,   
\begin{equation}
\label{highb}
    b\le b_{\rm max}=\sqrt{\frac{1}{\chi^2}-1}\ .
\end{equation}
\item As $\rho_+$ varies between $0$ and $\rho_b$, $b$ varies (monotonically) between $b_{\rm min}$ \eqref{lowb} and $b_{\rm max}$ \eqref{highb}. Note that $b_{\rm max}/b_{\rm min}=\sqrt{1+1/\chi}$ increases as $\chi$ decreases. See figure \ref{rhopbfig} for plots of (\ref{chib}) for several values of $\chi$.
\item According to \eqref{TM3}, the above variation of $\rho_+$ is related to the variation of the temperature from $0$ to $\infty$. The maximal value of $\rho_+$ is related to a maximal value of the energy, which we will compute below.
\end{itemize}

We now derive the energy of the caged black hole with these boundary conditions, following the discussion of section \ref{secthree}. In terms of $R_b$ (\ref{Rbb}) and $b$ (\ref{defrhobb}), the inverse temperature (\ref{TM3}) can be written as
\begin{equation}
\label{betaRbb}
\begin{split}
\beta={2\pi r_5\sqrt{(1+b^2)R_b^2-\rho_+^2}\sqrt{b^2\rho_+^2+(1+b^2)R_b^2}\over \rho_+(1+b^2)R_b}~.
\end{split}
\end{equation}
Using (\ref{chib}) to eliminate $\rho_+$ (for the case of fixed $\chi$), (\ref{betaRbb}) further reduces to
\begin{equation}
\begin{split}
\beta={2\pi r_5\chi(1+b^2)\sqrt{1-(1+b^2)\chi^2}\over \sqrt{\big((1+b^2)\chi\big)^2-1}}~.
\end{split}
\end{equation}
To evaluate the deformed energy for a black hole with horizon $\rho_+$, we use the thermodynamic relation $d E/d S=\beta^{-1}$. This gives the differential equation
\begin{equation}
\label{dEdba}
    \frac{dE}{db}=\frac{dE}{dS}\frac{dS}{db}=\frac{\pi}{2G_3\beta}\frac{d\rho_+}{db}={R_b\over 4 r_5G_3\chi}{1+(1+b^2)^2(2b^2-1)\chi^2\over b^2(\sqrt{1+b^2})^3\sqrt{1-(1+b^2)\chi^2}}\ .
\end{equation}
To obtain the second equality, we used $S={\pi\rho_+\over 2G_3}$, \eqref{Sbh}. 
Solving equation \eqref{dEdba}, we obtain
\begin{equation}
\label{Eb}
\begin{split}
    E={R_b\over 4 r_5G_3\chi}\left(2-\chi-\frac{1+2b^2}{b}\sqrt{\frac{1}{1+b^2}-\chi^2}\right)\ ,
\end{split}
\end{equation}
where the integration constant is picked such that when $\rho_+=0$, or $b^2=1/\chi-1$, \eqref{lowb}, $E=0$. This corresponds to tuning the cosmological constant of the deformed boundary theory to zero. 

Using the relation between the undeformed energy $E(0)$ and $\rho_+$, (\ref{Sfixed}) with (\ref{Sbh}), we can write (\ref{chib}) as 
\begin{equation}
\label{E0b}
    E(0)=\frac{R_b}{8r_5G_3}\frac{1+b^2}{b^2}\left((1+b^2)^2\chi^2-1\right)\ .
\end{equation}
Equations (\ref{Eb}) and (\ref{E0b}) should be understood as a parametric representation\footnote{Note that $E(0)$ \eqref{E0b} is a monotonically increasing function of $b$.} for $E$ as a function of $E(0)$. By multiplying these equations by $R_b$ we see that this dependence involves two dimensionless parameters, $r_5G_3/R_b^2$, and $\chi$. The former is familiar to us from section \ref{secthree} \eqref{lambdagravity}, while the latter is new -- it was fixed to $\chi=1$ in that section. It is natural to think of these two parameters as couplings in a deformed CFT. These couplings are irrelevant in the RG sense. For $r_5G_3/R_b^2$, this is due to similar considerations as in the previous sections, while for $\chi$ it can be seen as follows.

As explained above, the limit $b\to b_{\rm min}$ corresponds to $E(0)\to 0$ \eqref{E0b}. One can see from \eqref{Eb} that the energy $E$ behaves in this limit as 
\begin{equation}
\label{EE0}
    E=E(0)-\frac{2r_5G_3}{R_b}(1-2\chi)E(0)^2+O(E(0)^3)~.
\end{equation}
In particular, the coefficient of the linear term in \eqref{EE0} does not depend on $\chi$. Thus, the $\chi$ deformation does not change the low energy spectrum\footnote{Here, by low energy we mean energy much smaller than $R_b/(r_5G_3)$. Since we are discussing black holes, the states still satisfy $R_b E(0)\gg c_{\rm st}$.} -- its effect grows with the (undeformed) energy. This is of course the hallmark of an irrelevant deformation. 

\begin{figure}
	\centering
\includegraphics[scale=0.7]{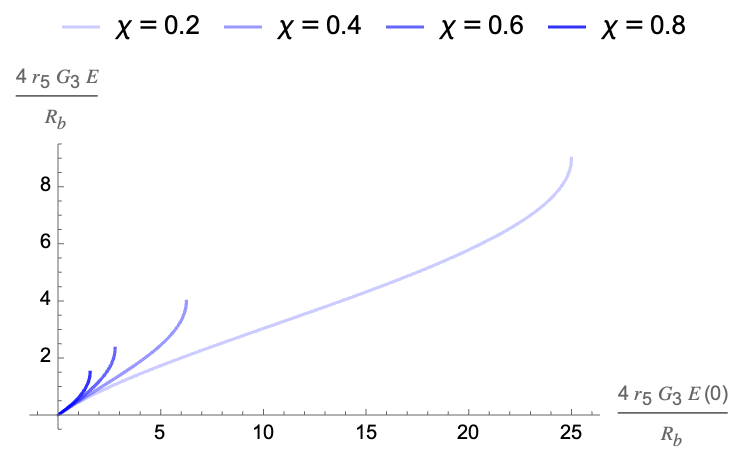}
\caption{\label{ESM3fig} $E$ (\ref{Eb}) as a function of $E(0)$ (\ref{E0b}) with fixed $\chi$.}
\end{figure}

In figure \ref{ESM3fig}, we plot $E$ as a function of $E(0)$ for several values of $\chi$. We see that $E$ increases monotonically with $E(0)$, and its low energy behavior is compatible with \eqref{EE0}. Besides, we can see from (\ref{Eb}) that as $b$ varies from $b_{\rm min}$ to $b_{\rm max}$, $E$ varies from 0 to 
\begin{equation}
\label{Emaxaa}
    E_{\rm max}={R_b\left(2-\chi\right)\over 4 r_5G_3\chi}\ .
\end{equation}
For $\chi=1$, \eqref{Emaxaa} reproduces the result of section \ref{secthree}, \eqref{Emaxa}. As $\chi$ decreases, the maximal energy increases, and eventually, as $\chi\to0$, it diverges, in agreement with the discussion of section \ref{sectwo} (where there is no maximal energy). Corresponding to $E_{\rm max}$, there is a maximal entropy, $S_{\rm max}=S(E_{\rm max})$. It is given by 
\begin{equation}
\label{Smaxchi}
    S_{\rm max}={\pi\rho_b(\rho_+=\rho_b)\over 2G_3}=\frac{\pi R_b}{2G_3\chi }\ .
\end{equation}
As mentioned above, the theory of section \ref{secthree} is obtained in the limit $\chi\to 1$ (which implies $b\to 0$, \eqref{lowb}, \eqref{highb}). Taking the limit with $E(0)$ fixed, it follows from (\ref{E0b}) that in the limit,
\begin{equation}
\label{bchilim}
    b=\sqrt{\frac{1-\chi}{1-\frac{4r_5G_3E(0)}{R_b}}}\ .
\end{equation}
Plugging \eqref{bchilim} in (\ref{Eb}), and taking the limit $\chi\to 1$, gives
\begin{equation}
\begin{split}
    E={R_b\over 4 r_5G_3}\left(1-\sqrt{1-\frac{8r_5G_3E(0)}{R_b}}\right)\ ,
\end{split}
\end{equation}
which is consistent with (\ref{Ebtzn0}), \eqref{E0Rb}.

Another limit we can consider is $\rho_b\gg R$, which leads to $b\to \infty$, \eqref{defrhobb}, and thus to the discussion of section \ref{sectwo}. In this limit, (\ref{E0b}) reduces to 
\begin{equation}
\label{rhopbgg1}
    E(0)=\frac{R_b}{8r_5G_3}\left((b^2\chi)^2-1\right)\ .
\end{equation}
Taking $b\to\infty$ with $E(0)$ fixed leads to $\chi\sim 1/b^2\to 0$. In this limit, (\ref{Eb})  becomes\footnote{Note that the subleading term of $\frac{1+2b^2}{b\sqrt{1+b^2}}$ is of order $1/b^4$, which is negligible compared to $\chi$.}
\begin{equation}
\begin{split}
    E={R_b\over 4 r_5G_3}\left(-1+\sqrt{1+\frac{8r_5G_3E(0)}{R_b}}\right)\ ,
\end{split}
\end{equation}
in agreement with (\ref{Ebh}). Thus, the energy formula \eqref{Eb}, \eqref{E0b} interpolates between the two theories described in sections \ref{sectwo}, \ref{secthree} as we vary $\chi$. This agrees with the qualitative picture, according to which as we vary the radial position of the cavity wall from the linear dilaton (large $\rho$) region to the AdS$_3$ (small $\rho$ one), the theory smoothly interpolates between those of sections \ref{sectwo} and \ref{secthree}, respectively.

\begin{figure}
	\centering
\includegraphics[scale=0.8]{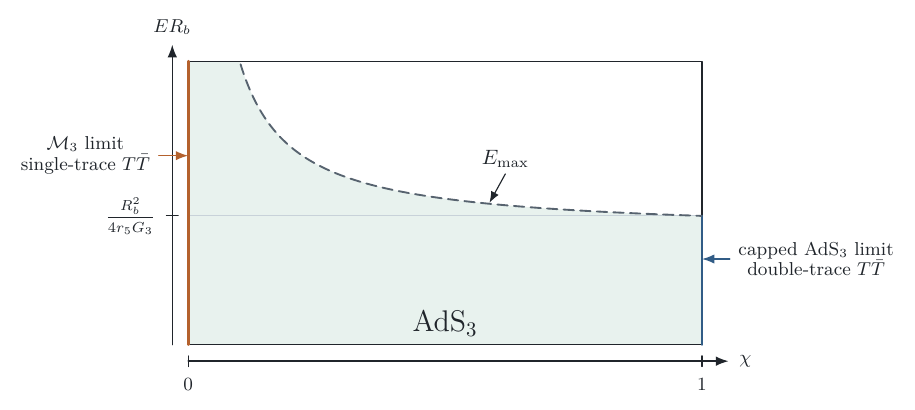}
\caption{\label{Echifig} The capped $\mathcal M_3$ theory simplifies in the limits $\chi\to 0,1$, and $E\ll R_b/r_5G_3$. The dashed line corresponds to \eqref{Emaxaa}.}
\end{figure}

The results of our analysis in this subsection are pictorially summarized in figure \ref{Echifig}. As we see in the figure, the black hole spectrum of gravity in capped ${\cal M}_3$, that corresponds to the shaded region below the curve \eqref{Emaxaa}, is controlled by an energy scale, $R_b/r_5G_3$, and a dimensionless parameter $\chi$, that measures the string coupling at the wall of the radial cavity. Each point in the shaded region in figure \ref{Echifig} is characterized by an entropy, or equivalently an undeformed energy $E(0)$, via equations \eqref{Eb}, \eqref{E0b}. For small $E$, the relation between the two is given in \eqref{EE0}. 
As $\chi\to 0,1$, the theory approaches the single and double trace $T\bar T$ deformed CFT, respectively. This is exhibited in the figure by the red and blue vertical lines. At low energies it  approaches the boundary CFT (for all $\chi$). 

For general values of the parameters and energy, it is a genuinely new theory. It is natural to ask what that theory is, i.e. what is the boundary interpretation of the spectrum \eqref{Eb}, \eqref{E0b} for general values of the parameters. We will leave a full understanding of this question to future work. However, here we would like to point out that the quadratic term in \eqref{EE0} has a natural interpretation in the boundary theory. Power counting suggests that it is due to a perturbation of the boundary CFT by a dimension four (more precisely $(2,2)$) operator. We saw that for $\chi=1$, this operator is the double trace $T\bar T$ operator, while for $\chi=0$, it is the single trace $T\bar T$ one. Thus, it is natural to define two couplings:
\begin{equation}
\label{interpr}
 \lambda_t= -{4r_5G_3\over R_b^2}\chi~, \qquad\lambda_d={4r_5G_3\over R_b^2}(1-\chi)~.
\end{equation}
Note that this definition is compatible with the limits $\chi\to 0,1$, where it reduces to \eqref{deflambda}, and  \eqref{lambdagravity}, respectively. 

The quadratic term in \eqref{EE0} looks like that in a $T\bar T$ deformed CFT with coupling $\lambda_d+\lambda_t={4r_5G_3\over R_b^2}(1-2\chi)$. This is consistent with the assumption that the leading deformation of the CFT is the combination \eqref{interpr} of single and double trace $T\bar T$. Of course, the higher order terms in the expansion \eqref{EE0} do not agree with that interpretation, which implies that some higher (than four) dimension   perturbations of the CFT are present as well. It would be interesting to find what they are, and rederive the spectrum \eqref{Eb}, \eqref{E0b} from the boundary perspective.

\subsection{Generalized Robin boundary conditions}

In the previous subsection, we studied the theory obtained by imposing Dirichlet boundary conditions on the dilaton $\Phi$ at the boundary of the radial cavity. In this subsection, we discuss a large class of more general boundary conditions for the dilaton.  We will again focus on the non-rotating case $\mu=0$. 

The boundary conditions we will impose are
\begin{equation}
\label{NPhi}
\partial_n\Phi|_{b}=F(\Phi_b)\ .
\end{equation}
Here, $\partial_n\Phi|_{b}$ denotes the normal derivative at the boundary, which is reparametrization invariant. In terms of the radial coordinate $\rho$ in (\ref{metric}), it takes the form
\begin{equation}
\label{NPhia}
\partial_n\Phi|_{b} = N(\rho_b)\Phi'(\rho_b)=-\frac{\sqrt{\rho_b^2-\rho_+^2}}{r_5}\frac{\rho_b}{\rho_b^2+R^2}\ .
\end{equation}
$F(\Phi_b)$ in (\ref{NPhi}) is a  function of the dilaton, which at this point in the discussion is arbitrary. The standard Robin boundary condition corresponds to 
\begin{equation}
F(\Phi_b)=A\,\Phi_b-B\ ,
\end{equation}
where $A$ and $B$ are constants. In the limit $A,B\to\infty$, with $A/B$ held fixed, (\ref{NPhi}) reduces to the Dirichlet boundary condition studied in the previous section, with $\chi=\frac{p}{kv}e^{2B/A}$.

The arbitrary function $F(\Phi_b)$ that appears in the specification of the boundary condition \eqref{NPhi} can be thought of in the coordinates (\ref{metric}) as parameterizing the radial position of the boundary $\rho_b$ as a function of the position of the horizon, $\rho_+$. This can be seen as follows.  Given a function $\rho_b(\rho_+)$ and using \eqref{Rbb}, (\ref{Phib}), one can determine the dependence of $\rho_+$ on $\Phi_b$:
\begin{equation}
\label{rhopPhib}    e^{2\Phi_b}=\frac{kvR_b\sqrt{R_b^2\rho_b^2(\rho_+)+\rho_+^2\left(\rho_b^2(\rho_+)-R_b^2\right)}}{p\rho_b^3(\rho_+)}\ .
\end{equation}
Using this relation, the r.h.s. of (\ref{NPhia}) depends only on $\Phi_b$, allowing us to rewrite this equation in the form (\ref{NPhi}), with
\begin{equation}
\label{FPhib}
    F(\Phi_b)= -\frac{\sqrt{[\rho_b(\rho_+(\Phi_b))]^2-[\rho_+(\Phi_b)]^2}}{r_5}\frac{[\rho_b(\rho_+(\Phi_b))]^2-R_b^2}{[\rho_b(\rho_+(\Phi_b))]^3}\ .
\end{equation}
Here, $\rho_b(\rho_+(\Phi_b))$ denotes a composite function of $\Phi_b$, with $\rho_+(\Phi_b)$ determined by (\ref{rhopPhib}).\footnote{
As mentioned earlier, the above discussion trivializes in the AdS$_3$ limit. Indeed, if $\rho_b(\rho_+)\ll R$, the dilaton $\Phi$ is fixed everywhere, to the value \eqref{btzdilaton}. Thus, the left hand side of \eqref{NPhi} vanishes, and one must have $F(\Phi_b)=0$. In this case, the discussion reduces to that of section \ref{secthree}.} 

A simple example of the boundary condition \eqref{NPhi} is the case where $\rho_b$ is a constant independent of $\rho_+$, analogous to the setup in section \ref{secthree}. We will analyze this case in detail in section \ref{secfixrhob}. For now, let us determine the corresponding function $F(\Phi_b)$. Combining (\ref{Rbb}) and (\ref{Phib}) gives the following dependence of $\rho_+$ on $\Phi_b$:
\begin{equation}
    \rho_+(\Phi_b)=\frac{R_b\rho_b}{\sqrt{\rho_b^2-R_b^2}}\sqrt{\left(\frac{pe^{2\Phi_b}\rho_b^2}{kvR_b^2}\right)^2-1}\ .
\end{equation}
Substituting it into (\ref{FPhib}), we obtain
\begin{equation}
\label{Frhob}
    F(\Phi_b)= -\sqrt{\frac{R_b^2}{\rho_b^2}-\left(\frac{pe^{2\Phi_b}}{kv}\right)^2}\frac{\sqrt{\rho_b^2-R_b^2}}{r_5R_b}\ .
\end{equation}
The AdS limit can be recovered by taking $\rho_b\to R_b$. In this limit, $F(\Phi_b)=0$, in agreement with the general discussion above.

\subsection{Single $+$ double trace $T\bar T$}\label{sec43}

As reviewed in section \ref{sectwo}, single trace $T\bar T$ deformed CFT is not completely understood. For a low energy observer, it looks like a symmetric product of CFT's of central charge $6k$, perturbed by a $\mathbb{Z}_2$ twisted modulus, however this description fails for high energy states of the sort we are discussing in this paper. Therefore, at first sight it seems unclear what happens when we add to single  trace $T\bar T$ deformed CFT a double trace perturbation. 

What makes this problem tractable in our setting is the following. A basic feature of the $T\bar T$ deformation of a general QFT is that given a state in the QFT on a circle, the effect of the $T\bar T$ deformation on the energy of that state is known (see e.g. the discussion around eq. (5.3) in \cite{Smirnov:2016lqw}). Thus, if we take the undeformed theory to be single trace $T\bar T$ deformed CFT, and the undeformed state to be the one corresponding to a black hole in ${\cal M}_3$, we can analyze the effect of the (double trace) $T\bar T$ deformation on it by using the above results. 

The ``undeformed'' states that correspond to black holes in ${\cal M}_3$ were discussed in section \ref{sectwo}, and were shown to have the spectrum \eqref{Elambda}. Applying a $T\bar T$ deformation with dimensionless coupling $\lambda_t$ to this spectrum leads to the same spectrum, but with the coupling $\lambda_d$ in \eqref{Elambda} replaced by $\lambda=\lambda_d+\lambda_t$. Thus, in this subsection we will ask the question whether we can choose the function $F(\Phi_b)$ such that the resulting spectrum has this form. 

Note that the black holes are only sensitive to $\lambda$, and not $\lambda_d$ and $\lambda_t$ separately. However, other features of the theory, like the spectrum of perturbative string states, are sensitive to the two couplings, and not just their sum, since changing the two couplings while keeping their sum fixed changes the local geometry and position of the cavity wall. 

The starting point of our analysis, eq. \eqref{Elambda} with $\lambda_d\to\lambda=\lambda_d+\lambda_t$, can be written in terms of the entropy of the state, $S$, rather than its undeformed energy $E(0)$, as 
\begin{equation}
\label{fourse}
R_bE(S)=\frac{1}{\lambda}\left(-1+\sqrt{1+\lambda \frac{G_3S^2}{\pi^2r_5}}\right)\ .
\end{equation}
For $\lambda>0$ ($\lambda<0$), the deformation is qualitatively similar to a single-trace (double-trace) $T\bar T$ deformation. In particular, for $\lambda>0$ the energy is not bounded from above, while for $\lambda<0$ there is a maximal energy, $E_{\rm max}=-1/(\lambda R_b)$.  

\begin{figure}
	\centering
    \subfigure[]{
	\begin{minipage}[t]{0.45\linewidth}
	\centering
	\includegraphics[width=2.8in]{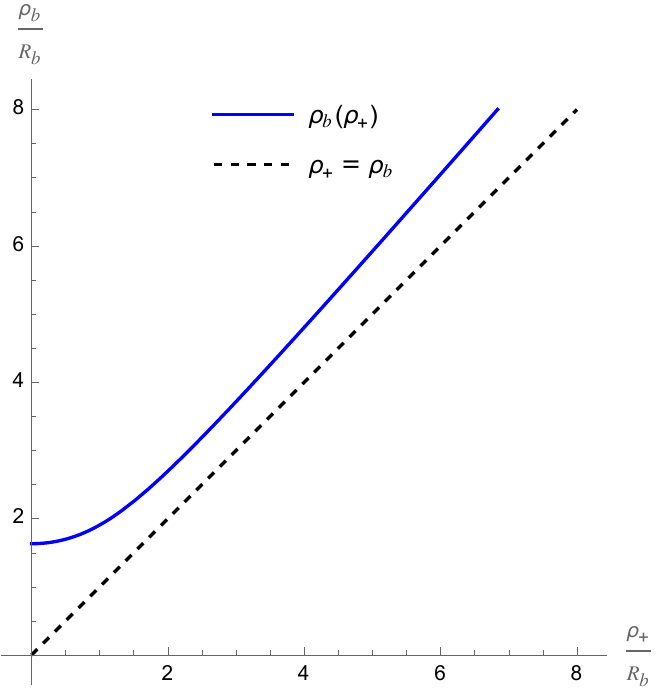}\label{rhobrhoplamp}
	\end{minipage}}
	\subfigure[]{
	\begin{minipage}[t]{0.45\linewidth}
	\centering
	\includegraphics[width=2.8in]{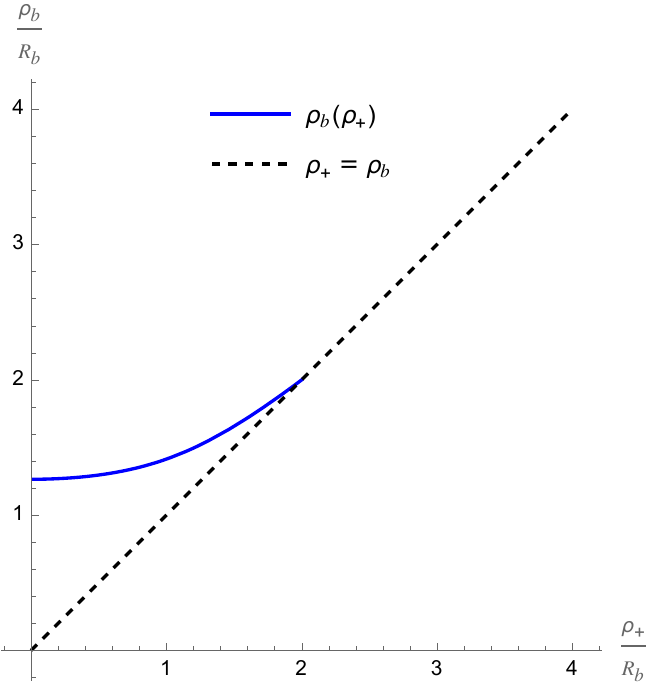}\label{rhobrhoplamn}
	\end{minipage}}
	\centering
\caption{\label{rhobrhopfig}$\rho_b$ as a function of $\rho_+$, (\ref{rhoprhobttbar}), for $T\bar T$ deformation parameter (a) $\lambda=r_5G_3/R_b^2$, (b) $\lambda=-r_5G_3/R_b^2$.}
\end{figure}

To determine the corresponding function $\rho_b(\rho_+)$, and the corresponding $F(\Phi_b)$ \eqref{NPhi}, we proceed as follows. Using \eqref{fourse}, we can express the inverse temperature as
\begin{equation}
\label{betaTTbar}
\beta(S)=\left(\frac{\partial E}{\partial S}\right)^{-1}=\frac{\pi^2r_5R_b\sqrt{1+\lambda \frac{G_3S^2}{\pi^2r_5}}}{G_3S}=\frac{2\pi r_5R_b\sqrt{1+\lambda \frac{\rho_+^2}{4r_5G_3}}}{\rho_+}\ ~.
\end{equation}
Further using (\ref{Rbb}) and (\ref{TM3}) yields
\begin{equation}
\label{rhoprhobttbar}
\rho_+=\frac{\rho_b\sqrt{\rho_b^2-\frac{\lambda}{4r_5 G_3}R_b^2\rho_b^2-2R_b^2}}{\sqrt{\rho_b^2-R_b^2}}\ .
\end{equation}
In figure \ref{rhobrhopfig}, we plot $\rho_b(\rho_+)$ for positive and negative $\lambda$. We see that for $0\le\lambda\le \lambda_*$, where 
\begin{equation}
\label{lamstar}
\lambda_*={4r_5G_3\over R_b^2}~,
\end{equation}
$\rho_b>\rho_+$ for all $\rho_+$ (see figure \ref{rhobrhoplamp}). This is responsible for the fact that there is no maximal energy in this case. At large $\rho_+$, $\rho_b$ grows linearly, with a slope greater than 1. Specifically, it follows from (\ref{rhoprhobttbar}) that for $0\le\lambda<\lambda_*$, the large $\rho_+$ behavior is
\begin{equation}
\label{nowall}
    \rho_b\sim \frac{\rho_+}{1-\frac{\lambda}{\lambda_*}}\ .
\end{equation}
As $\lambda\to\lambda_*$, both the slope and the intercept $\rho_b(0)$ diverge, and for $\lambda>\lambda_*$, eq. \eqref{rhoprhobttbar} does not have real solutions (with $\rho_b>R_b$ \eqref{Rbb}). The behavior \eqref{nowall} is interesting -- the construction involves a radial cavity, but we never reach its wall, since as the energy of states increases, it recedes to infinity. 

Similarly, for negative $\lambda$, the physical region is $-\lambda_*\le\lambda\le 0$. This follows from the requirement that the solid blue curve in figure \ref{rhobrhoplamn} remains in the physical region $\rho_b>R_b$. Thus, we conclude that $\lambda$ in our analysis above, \eqref{fourse}, is restricted to the region $|\lambda|\le\lambda_*$. One can think of this constraint in a similar way to equation \eqref{interpr}. We write  $\lambda_t=-y\lambda_*$ and $\lambda_d=(1-y)\lambda_*$, where $y$ is a parameter that lives in the interval $0\le y\le 1$. In this parametrization, $\lambda=\lambda_t+\lambda_d=(1-2y)\lambda_*$ ranges from $-\lambda_*$ to $+\lambda_*$, and the energy formula \eqref{fourse} is exactly that of single $+$ double trace deformed CFT. 

Substituting (\ref{rhoprhobttbar}) into (\ref{Phib}) determines $\rho_b(\Phi_b)$:
\begin{equation}
\label{rhobPhibttbar}
e^{2\Phi_b}=\frac{kvR_b\sqrt{\rho_b^2-\frac{\lambda}{4r_5G_3}R_b^2\rho_b^2-R_b^2}}{p\rho_b^2}\ .
\end{equation}
Using (\ref{rhoprhobttbar}) and (\ref{rhobPhibttbar}), we can eliminate $\rho_+$ and $\rho_b$ on the r.h.s. of (\ref{NPhia}), and thus write it as a function of $\Phi_b$, which takes the form
\begin{equation}
\label{sindoub}
F(\Phi_b)=-\frac{1}{r_5}\sqrt{\left(\frac{pe^{2\Phi_b}}{kv}\right)^2+\frac{\lambda R_b^2}{4r_5G_3}}\ .
\end{equation}
The square root in \eqref{sindoub} is very similar to the one in (\ref{Frhob}). The $e^{4\Phi_b}$ term in the former is equal, up to a sign, to that in the latter. The second term in the square root in \eqref{sindoub} is closely related to that in (\ref{Frhob}) as well. To see that, we note that here $\rho_b$ depends on the state (via \eqref{rhoprhobttbar}), but for negative $\lambda$ it has a maximal value, obtained by taking $\rho_+=\rho_b$ (see figure \ref{rhobrhoplamn}). Denoting this value by $\rho_b^{\rm (max)}$, the second term in the square root \eqref{sindoub} can be written as $-(R_b/\rho_b^{\rm (max)})^2$. Thus, in this interpretation, the arguments of the square roots in (\ref{Frhob}) and \eqref{sindoub} are related by an overall minus sign. Of course, in (\ref{Frhob}) there is an additional factor of $\sqrt{\rho_b^2-R_b^2}/R_b$ which is absent in \eqref{sindoub}. 

An interesting limit of the discussion of this subsection is $\lambda\to 0$. The curve $\rho_b(\rho_+)$ has in this case the property that $\rho_b(\rho_+)$ (solid blue line in figure \ref{rhobrhopfig}) approaches the line $\rho_b=\rho_+$ (dashed line) as $\rho_+\to\infty$. The function $F(\Phi_b)$ \eqref{sindoub} simplifies in this limit, becoming a pure exponential. The energy formula \eqref{fourse} becomes trivial in the limit $\lambda=\lambda_d+\lambda_t\to 0$, as the deformed and undeformed energies coincide. Thus, it seems that the effects on the spectrum of black holes of generalizing from AdS$_3$ to ${\cal M}_3$ and introducing the cavity wall cancel each other, and the high energy of density of states of the resulting theory exhibits Cardy behavior. It is natural to ask whether this theory is a CFT. To see that it isn't, one can look at the behavior of perturbative string states, in particular the long strings that belong to the principal continuous series representations in the original AdS$_3$ theory, and are described in the boundary CFT by the symmetric product $(M_{6k})^N/S_N$. One can show that their energies experience a non-zero $T\bar T$ deformation, and  their correlation functions differ from those of a CFT$_2$.~\footnote{For the energies, this follows from eq. (42) in \cite{Chakraborty:2024ugc}; for correlation functions it follows from \cite{Cui:2023jrb,Giveon:2023gzh}.}

Thus, the theory described in this subsection has the interesting property that it is non-local, but at $\lambda=0$ has the high energy density of states of a two dimensional CFT.\footnote{We thank N. Itzhaki for a discussion on this.} It lies at the intersection between theories with a Hagedorn spectrum ($\lambda>0$), and theories with a finite entropy ($\lambda<0$).

\subsection{Fixed $\rho_b$}\label{secfixrhob}

In this subsection, we study the theory in which the cavity wall is fixed in the $\rho$ coordinates, i.e $\rho_b$ is independent of $\rho_+$. As we discussed before, the reparameterization invariant way of describing it is as a special case of (\ref{NPhi}), with the function $F$ given by (\ref{Frhob}). 

Equation (\ref{Rbb}) implies that in this case both $R$ and $\rho_b$ are fixed features of the theory (i.e. do not depend on the state). Thus, the parameter $b$, (\ref{defrhobb}), is also a feature of the deformed theory. As before, varying it between zero and infinity interpolates between the limit where the boundary is deep inside the AdS$_3$ region and the one where it is deep inside the linear dilaton one. 

Taking $b$ and $R_b$ to be the independent parameters, $R$ can be expressed in terms of them as
\begin{equation}
R=\frac{\sqrt{1+b^2}R_b}{b}\ .
\end{equation}
For small $b$, this relation takes the form $R_b=bR=\rho_b$, in agreement with the discussion of section \ref{secthree}, \eqref{Rb}, while for large $b$ we get $R_b=R$,  in agreement with the discussion of section \ref{sectwo}.

Following the by now familiar route, we rewrite the inverse temperature (\ref{TM3}) in terms of the constants $b$ and $R_b$, and the entropy $S=\frac{\pi\rho_+}{2G_3}$, 
\begin{equation}
\label{TofS0M3}
\begin{split}
\beta(S)={r_5\sqrt{(2bG_3S)^2+(1+b^2)(\pi R_b)^2}\sqrt{(1+b^2)(\pi R_b)^2-(2G_3S)^2}\over G_3S R_b(1+b^2) }~.
\end{split}
\end{equation}
As a check, \eqref{TofS0M3} reduces to \eqref{TofS0} in the limit $b\to 0$.

Integrating \eqref{TofS0M3} over $S$ gives the energy, in a form analogous to \eqref{Ebtz},
\begin{equation}
\label{Edeformed}
E=\frac{(1+b^2)R_b}{4br_5G_3}\left(\sin^{-1}\frac{\sqrt{8b^2r_5^2G_3M+(1+b^2)R_b^2}}{(1+b^2)R_b}-\sin^{-1}\frac{1}{\sqrt{1+b^2}}\right) .
\end{equation}
After some manipulation, \eqref{Edeformed} can be shown to take the form
\begin{equation}
\label{abag}
   2m(1+\sin\gamma)+\sin\gamma=\sin(\alpha+\gamma)~,
\end{equation}
where 
\begin{equation}
\label{abaf}
\cos\gamma=\frac{2b}{1+b^2}~,\;\;\;\sin\gamma=\frac{1-b^2}{1+b^2}~,
\end{equation}
\begin{equation}
\label{defmm}
m={4b^2r_5^2G_3M\over (1+b^2) R_b^2}={2r_5G_3E(0)\over R_b}(1-\sin\gamma)~,
\end{equation}
\begin{equation}
\label{abbg}
\alpha=\frac{ 8br_5G_3 E}{(1+b^2)R_b}=\frac{ 4r_5G_3 }{R_b}E\cos\gamma~.
\end{equation}
The spectrum \eqref{abag} -- \eqref{abbg} can be thought of from the boundary point of view as the spectrum of a theory with two couplings. One way to parametrize these couplings is to define\footnote{Note the intriguing similarity to \eqref{interpr}, that we obtained for  Dirichlet b.c.'s. The role of $\chi$ there is played by $\sin^2\theta$ here.}
\be\label{definee}
\begin{split}
&\lambda_d={4r_5G_3\over R_b^2}\cos^2\theta~,\\ 
&\lambda_t=-{4r_5G_3\over R_b^2}\sin^2\theta~,
\end{split}
\ee
where $\theta$ satisfies
\begin{equation}
\label{deftheta}
\sin\theta={1\over\sqrt{1+b^2}}~,\;\;\;\cos\theta={b\over\sqrt{1+b^2}}~.
\end{equation}
As is familiar from the literature on $T\bar T$ deformed CFT, one can think of the dimensionless couplings \eqref{definee} as coming from the dimensionful couplings $\lambda_iR_b^2$, which correspond to irrelevant deformations of the original CFT$_2$, evaluated at the KK scale $R_b$.

The dimensionless parameter $b$  \eqref{defrhobb} can be thought of as the ratio of the above couplings, $b=\cot\theta=\sqrt{\lambda_d/|\lambda_t|}$. As $\theta$ varies between $0$ and $\pi/2$, $b$ varies between infinity and zero. Note also that $\theta$ \eqref{deftheta} is related to $\gamma$ \eqref{abaf} via the relation $\gamma=2\theta-{\pi\over2}$, so $\gamma$ varies between $-{\pi\over2}$ and $\pi\over2$. And, one can write \eqref{Rbb} in terms of $\theta$ as $R_b=R\cos\theta$. 

A useful property of the definitions \eqref{definee}, \eqref{deftheta} is that the spectrum \eqref{abag} -- \eqref{abbg} satisfies the simple relation,
\begin{equation}
\label{dEdE}
    \frac{dE(0,0)}{d E(\lambda_d,\lambda_t)}=\sqrt{1+2\lambda_dR_bE(0,0)}\sqrt{1+2\lambda_tR_bE(0,0)}\ .
\end{equation}
This relation is useful for studying the limits where one of the two couplings \eqref{definee} goes to zero. It is easy to see that when $\lambda_t\to 0$, \eqref{dEdE} goes to \eqref{Elambda} (with $P=0$), while for $\lambda_d\to 0$, it approaches \eqref{Ebndry} (again, with $P=0$). 

Another interesting limit of \eqref{dEdE} is small $\lambda_i$ $(i=t,d)$, and arbitrary $b$. Expanding to first order in $\lambda_i$ we see that, to that order, the deformed energies behave like those of a $T\bar T$ deformed CFT with dimensionless coupling $\lambda_t+\lambda_d$, as expected. However, at higher order in $\lambda_i$, this relation breaks down, again as expected (from the form of the spectrum \eqref{abag}). 

We finish this subsection by pointing out a few additional properties of the spectrum \eqref{abag} -- \eqref{abbg}. 

The line $b=1$ in the two dimensional coupling space labeled by $(\lambda_d,\lambda_t)$ corresponds  to $\theta=\pi/4$, i.e. $\lambda_d+\lambda_t=0$, \eqref{definee}. The spectrum \eqref{abag} -- \eqref{abbg} simplifies further in this case. It takes the form 
\begin{equation}
\label{bbone}
\sin\left(\lambda R_b E(\lambda)\right)=\lambda R_bE(0)~,
\end{equation}
where $\lambda\equiv\lambda_d-\lambda_t={4r_5G_3\over R_b^2}$.
It would be interesting to understand what is the theory with this intriguingly simple spectrum, or more generally the one with the spectrum \eqref{abag} -- \eqref{abbg}.

We mentioned in section \ref{secthree} that the entropy of $T\bar T$ deformed CFT satisfies the constraint \eqref{Sfixed}. This property is valid in the more general cases discussed in this section, where both $\lambda_d$ and $\lambda_t$ are turned on, essentially for the same reason. However, unlike in section \ref{secthree}, here we do not expect this property to be valid beyond the high energy spectrum. The reason is that as was shown in \cite{Aharony:2018bad}, the $T\bar T$ deformed spectrum is the unique one that satisfies \eqref{Sfixed} for all states, and our spectrum is different. This observation is not particularly surprising, since it is already true for single trace $T\bar T$, that is described in section \ref{sectwo}.

\begin{figure}
	\centering
\includegraphics[scale=0.8]{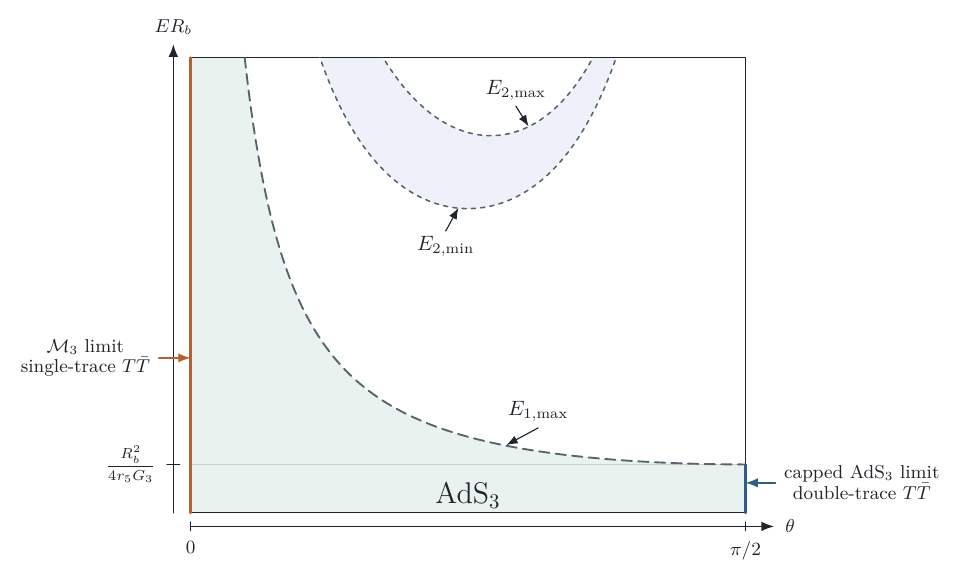}
\caption{\label{Ethetafig} The capped $\mathcal M_3$ theory has a band structure discussed in the text. The black holes are described by the shaded green region, and $E_{1,\rm max}$ is given by \eqref{maxen}. The first excited band is described by the shaded purple region. It is shifted by the amount \eqref{bands}, with $N=1$ w.r.t. the black hole band.}
\end{figure}

Looking back at \eqref{abag}, we note that it has two additional interesting properties. The first is that as one varies $E(0)$ from zero, the deformed energy $E(\lambda_t,\lambda_d)$ increases, until one gets to the point where $\sin(\alpha+\gamma)=1$, where the deformed energy is largest. Thus, the maximal deformed energy is the one for which $\alpha+\gamma=\pi/2$,
\begin{equation}
\label{maxen}
E_{\rm max}=\frac{R_b(\pi-2\theta)}{4r_5G_3\sin2\theta}~,
\end{equation}
where we used the relation between $\gamma$ and $\theta$ mentioned above. It is denoted by $E_{1,\rm max}$ in figure \ref{Ethetafig}. In the bulk, the maximal energy corresponds to a black hole whose horizon coincides with the boundary of the cavity, $\rho_+=\rho_b$. This is easily checked by plugging  $\rho_+=\rho_b$ in \eqref{Edeformed} and expressing the result  in terms of $\theta$. Moreover, the requirement that $\beta$ be real implies $\rho_+\leq\rho_b$, (\ref{TofS0M3}). Hence, integrating over $S$ up to the corresponding upper bound indeed yields the maximal energy. The maximal entropy is given by
\begin{equation}
    S_{\rm max}={\pi\rho_b\over 2G_3}=\frac{\pi R_b}{2 G_3\sin\theta}\ ,
\end{equation}
which has the same form as (\ref{Smaxchi}), with $\chi$ there replaced by $\sin\theta$ here.

Equation \eqref{abag} gives the energies of the states that correspond in the bulk to black holes in ${\cal M}_3$. A second interesting feature of this equation is that it has additional solutions that do not have this interpretation. Indeed, for a given undeformed energy $E(0)$, the l.h.s. of \eqref{abag} is fixed, but the r.h.s. is a periodic function of the deformed energy $E=E(\lambda_d,\lambda_t)$. Thus, if we replace $E$ by 
\begin{equation}
\label{bands}
E\to E+\frac{2\pi NR_b}{4r_5G_3\cos\gamma}~,
\end{equation}
for any integer $N$, the equation is still satisfied. Note that the energy difference between adjacent energy levels goes to infinity as $\gamma\to\mp{\pi\over2}$. The two edge cases correspond to $\theta=0,{\pi\over2}$, which were the topics of sections \ref{sectwo} and \ref{secthree}, respectively. In these cases, there is no repeating pattern such as \eqref{bands}, in agreement with the literature on the subject. We plot the first two energy bands in figure \ref{Ethetafig}. 

An interesting feature of \eqref{bands} is that the energy difference between adjacent bands, $\Delta E=E_{N+1}-E_N$, is non-perturbative in the couplings \eqref{definee}. Indeed, one can write it as 
\begin{equation}
\label{nonpert}
\Delta E=\frac{\pi}{R_b\sqrt{\lambda_d|\lambda_t|}}~.
\end{equation}
Thus, the states with $N\not=0$, if they exist, disappear when we study the theory perturbatively in the couplings. It is an interesting question what these potential new states correspond to in the bulk. Since their energy differs from that of black holes by an amount that goes like $1/g^2$, they must correspond to new geometries. It would be nice to determine if these states are really there, and if so to identify them.

\section{Discussion}
\label{secfive}

Single and double trace $T\bar T$ deformed CFT  in string theory give rise to conceptually interesting and practically useful insights into the holographic duality between three dimensional bulk backgrounds that are not asymptotically AdS$_3$, and boundary theories that do not approach RG fixed points in the UV. This motivates the search for generalizations of the framework to a larger class of spacetimes. In this paper, we took a step in this direction.  

We did this by combining the elements that enter the construction of single and double trace $T\bar T$ deformed CFT. The former corresponds in the bulk to generalizing the background from AdS$_3$ to ${\cal M}_3$, a $2+1$ dimensional spacetime that smoothly interpolates between a linear dilaton spacetime, $\mathbb{R}_\phi\times\mathbb{R}_t\times\mathbb{S}^1$, in the UV, and AdS$_3$ in the IR. The latter corresponds to placing a cutoff on the radial direction in AdS$_3$, and  considering the theory in the resulting spherical cavity. 

To combine the two, we studied string theory in a spherical cavity in ${\cal M}_3$. The double trace $T\bar T$ theory corresponds to a limit of this construction, in which the cavity is located deep inside the AdS$_3$ region of ${\cal M}_3$, but in general the cavity wall can be anywhere in the radial direction. When the wall is sent to infinity, the theory approaches the single trace $T\bar T$ one. 

Therefore, any such construction includes a parameter that corresponds to the radial position in ${\cal M}_3$ of the wall of the cavity relative to the radial position where the transition between the linear dilaton region and AdS$_3$ occurs. This parameter can be thought of as an extra coupling in the boundary theory. As one varies it, the construction continuously interpolates between the double trace and single trace $T\bar T$ theories. For general values of this coupling, it gives rise to a new theory. 

We found that in ${\cal M}_3$ one can impose a large class of boundary conditions on the dilaton, which lead to different theories. We studied black holes in the resulting backgrounds, and used thermodynamic considerations to compute their spectra in the radial cavity. The different constructions give rise to different spectra, which we examined in a few cases. In particular, for one choice of boundary conditions, we found that the boundary theory is a double trace $T\bar T$ deformation of single trace $T\bar T$ deformed CFT, with the two couplings alluded to above being the single and double trace $T\bar T$ couplings. Another natural example gives rise to a rather exotic spectrum, which contains an infinite number of energy bands, the lowest of which consists of black hole states. 

We showed that the freedom of imposing boundary conditions in the class we studied disappears in the limit in coupling space where the cavity is deep in the AdS$_3$ region. In other words, all these theories have the property that they reduce to the double trace $T\bar T$ construction of \cite{McGough:2016lol} in that limit. Similarly, in the region where the cavity recedes to infinity, they approach single trace $T\bar T$ deformed CFT. We checked that the examples we considered in detail indeed have this property. Another general property of our construction is that the leading correction to the CFT spectrum in the low energy expansion is always a combination of single and double trace $T\bar T$ deformations. The different boundary conditions correspond to different values of higher order irrelevant couplings.

Much remains to be understood about these theories. Perhaps the most important open problem is to understand these theories from the boundary point of view. The examples we considered in detail look like UV deformations of the CFT$_2$ dual to string theory on AdS$_3$, perturbed by an in general correlated infinite set of irrelevant operators. It is likely that one can describe these perturbations directly from the boundary point of view, and it would be very interesting to do so, since it would extend the success of the $T\bar T$ paradigm to a much larger set of theories. 

The class of boundary conditions we studied involves a single function of the boundary value of the dilaton, $F(\Phi_b)$ \eqref{NPhi}. It would be interesting to understand what is the significance of this fact, and in particular to understand how the qualitative structure of the theory depends on the choice of the function $F$. For example, in the special cases we studied, in some the construction led to a theory with a finite maximal energy and entropy, and in others those were infinite. It would be interesting to understand what features of $F$ control this behavior. 

As we saw in section \ref{secfour}, another way to parameterize the generalized Robin boundary conditions is in terms of the position of the cavity wall, $\rho_b(\rho_+)$. The qualitative structure of the theory depends on the behavior of this function. If $\rho_b(\rho_+)>\rho_+$ for all $\rho_+$, the theory does not have a maximal energy, and in general behaves similarly to that of section \ref{sectwo}, despite describing a cavity. In particular, generically it has a Hagedorn density of states at high energies.\footnote{If $\rho_b/\rho_+$ goes to infinity at large $\rho_+$, eq. \eqref{TM3} implies that $\beta_H=2\pi r_5$. If it goes to a constant larger than one, as in section \ref{sec43}, $\beta_H$ depends on this constant.} 
In the special subclass of theories in which $\rho_b(\rho_+)$ approaches $\rho_+$ (from above) at large $\rho_+$, the theory has a Cardy entropy at high energies. 

A second class of theories is one in which $\rho_b$ has the qualitative structure of figure \ref{rhobrhoplamn}. In this case, $\rho_b>\rho_+$ in a finite range of $\rho_+$, starting from the origin. The resulting theories generally have a maximal energy and finite entropy, and in that respect are similar to the theory of section \ref{secthree}. We have seen a couple of examples of such theories in section \ref{secfour}. One can also consider more exotic situations, where $\rho_b>\rho_+$ holds between two finite values of $\rho_+$, so that there is both a minimal and a maximal energy. More generally, one can consider situation where this equation is satisfied in a union of several line segments. We leave the study of such theories to future work.   

In \cite{Guica:2019nzm}, a different perspective on the construction of \cite{McGough:2016lol} is presented. In this description, the boundary conditions on the gravity fields are imposed on the boundary of AdS$_3$, and the radial cavity description is an effective description of these boundary conditions. It would be interesting to generalize their construction to our setting, where AdS$_3$ is replaced by ${\cal M}_3$, and in particular understand the role of the function $F(\Phi_b)$ in that language. 

In the discussion in section \ref{secfour} we restricted our attention to the zero momentum case. It would be interesting to generalize the discussion to non-zero momentum. One question that needs to be addressed to this end is the generalization of the boundary conditions we used to non-zero momentum. We leave a detailed study of this issue to future work. 

The starting point of our analysis was {\it positive} single trace $T\bar T$ deformed CFT. It would be interesting to generalize the discussion to the {\it negative} single trace $T\bar T$ theory. This theory was studied in \cite{Chakraborty:2020swe,Chakraborty:2023zdd,Giveon:2024sgz,Vainshtein:2026ixi}. It describes a spacetime that looks like AdS$_3$ in the IR, and has a spacelike singularity at a finite value of the radial coordinate. It provides an interesting generalization of our work, since it involves questions like what happens when the wall of the radial cavity approaches the singularity. This too will be left for future work.  

It would also be interesting to investigate further the theory studied in subsection \ref{secfixrhob}. The spectrum we found in that case exhibits hints of discretization of spacetime. This raises a number of questions, like if the spacetime is indeed discrete, what happens to energy and momentum conservation, what features of the boundary conditions are responsible for the discrete structure we observe, and how common is the appearance of this discrete structure in the space of $F(\Phi_b)$. 

Finally, our construction might be useful for studying various pathologies that were discussed in the literature for spacetimes with radial cavities. These pathologies include superluminal signal propagation, and the associated issues with causality \cite{McGough:2016lol}, dynamical instabilities \cite{Anninos:2023epi,Helton:2026zro}, and problems with defining the Euclidean path integral with these boundary conditions \cite{Witten:2018lgb,An:2025gvr}. In our models, the embedding of the radial cavity in ${\cal M}_3$ rather than AdS$_3$ gives rise to additional parameters, such as $\chi$ in section \ref{Dirichlet} and $\theta$ in section \ref{secfixrhob} that, for example, allow us to vary the maximal energy $E_{\rm max}$ for a fixed value of the gravitational parameters $(R_b, G_3, r_5)$, see figures \ref{Echifig} and \ref{Ethetafig}. Our construction also allows one to interpolate continuously between models with a maximal energy and finite entropy, and ones with a Hagedorn high energy density of states. This might help clarify the physical implications and potential resolutions of these pathologies.

\section*{Acknowledgements}

JC thanks Westlake University for hospitality during the conclusion of this work. DK thanks O. Aharony and N. Itzhaki for discussions. 
The work of AG was supported in part by the ISF (grant number 256/22). The work of JC and DK was supported in part by DOE grant DE-SC0009924.


\vskip 2cm

\bibliographystyle{JHEP}
\bibliography{ref}

\end{document}